\documentclass[12pt]{article}
\usepackage{latexsym}
\usepackage{amssymb}
\usepackage{graphicx}
\usepackage{multibox}
\usepackage{amsmath,amsfonts}

\topskip \textwidth 480pt

\def\*{\ast}

\def\ve{\varepsilon}

\def\be{\begin{equation}}
\def\ee{\end{equation}}
\def\bqn{\begin{eqnarray}}
\def\eqn{\end{eqnarray}}

\def\theequation{\thesection.\arabic{equation}}

\newsavebox{\ver}
\newsavebox{\verp}

\newsavebox{\gorp}
\newsavebox{\toch}

\newcommand{\bee}{\begin{eqnarray}}
\newcommand{\eee}{\end{eqnarray}}

\newcommand{\ups}{\upsilon}

\date{}
\begin{document}
\begin{titlepage}
\title{
\begin{flushright}
{\small NSF-KITP-09-38}\\
~\\
~\\
\end{flushright}
{\bf Simplifying superstring and D--brane actions in $AdS_4\times
CP^3$ superbackground}
\medskip
\medskip
\author{Pietro Antonio Grassi\footnote{\tt pgrassi@cern.ch}\,~$^\ddag$, Dmitri~Sorokin\footnote{\tt dmitri.sorokin@pd.infn.it, linus.wulff@pd.infn.it}~ and Linus Wulff $^{\dagger}$
~\\
~\\
{\it $^*$  DISTA, Universit\'a del Piemonte Orientale,} ~\\
 {\& \it  INFN, Gruppo
Collegato Sezione di Torino,} ~\\
{\it via
Bellini 25/g, 15100 Alessandria, Italia} ~\\
~\\
{\it $^\dagger$ Istituto Nazionale di Fisica Nucleare, Sezione di
Padova,}
~\\
{\it via F. Marzolo 8, 35131 Padova, Italia}
~\\
~\\
{\it $^\ddag$ Kavli Institute for Theoretical Physics,}
~\\
{\it University of California at Santa Barbara, CA 93106-4030, USA}}
}
\maketitle

\begin{abstract}
By making an appropriate choice for gauge fixing kappa--symmetry we
obtain a relatively simple form of the actions for a $D=11$
superparticle in $AdS_4\times S^7/Z_k$, and for a D0--brane,
fundamental string and D2--branes in the $AdS_4\times CP^3$
superbackground. They can be used to study various problems of
string theory and the $AdS_4/CFT_3$ correspondence, especially in
regions of the theory which are not reachable by the
$OSp(6|4)/U(3)\times SO(1,3)$ supercoset sigma--model. In
particular, we present a simple form of the gauge--fixed superstring
action in $AdS_4\times CP^3$ and briefly discuss issues of its
T--dualization.
\end{abstract}
~

\thispagestyle{empty}
\end{titlepage}
\tableofcontents
\section{Introduction}

Recent developments, initiated in
\cite{Bagger:2006sk,Gustavsson:2007vu}, which led to important
progress in understanding the holographic duality between $D=3$
superconformal theories and type IIA string/M--theory on $AdS_4$
have revived an interest in studying strings and branes in
supergravity backgrounds whose bosonic subspace is $AdS_4\times
M^{6}$ and $AdS_4\times M^{7}$, respectively, where $M^6$ is a
compactified manifold of $D=10$ type IIA supergravity and $M^7$ is
its Hopf fibration counterpart in $D=11$ supergravity (or
M--theory). Examples of interest include the supergravity solutions
with $M^6=CP^3$ and $M^7=S^7/Z_k$ (with an integer $k$ being the
Chern--Simons theory level) and their squashings.

In particular, the ${\cal N}=6$ Chern-Simons theory with the gauge
group $U(N)_k\times U(N)_{-k}$ \cite{Aharony:2008ug} has been
conjectured to describe, from the $CFT_3$ side, M--theory on $AdS_4
\times S^7/Z_k$. In the limit of the parameter space of the ABJM
theory in which the 't Hooft coupling $\lambda={N/k}$ is
$\lambda^{5/2}<<N^2$ and $k>>1$, the bulk description is given in
terms of perturbative type IIA string theory on the $AdS_4
\times CP^3$ background. To analyze this new type of holographic
correspondence from the bulk theory side, an explicit form of the
action for the superstring in $AdS_4 \times CP^3$ superspace is
required.

In contrast to \emph{e.g.} the case of type IIB string theory in
$AdS_5 \times S^5$ superspace which preserves the maximum
number of 32 supersymmetries and is thus described by the supercoset
$PSU(2,2|4)/SO(1,4)\times SO(5)$, the case of type IIA string theory
on $AdS_4 \times CP^3$ is more complicated since $AdS_4 \times CP^3$
preserves only 24 of 32 supersymmetries. As a consequence, the
complete type IIA superspace with 32 fermionic coordinates, that
solves the IIA supergravity constraints for the $AdS_4 \times CP^3$
vacuum solution, is not a coset superspace. This superspace has been
constructed in
\cite{Gomis:2008jt} by dimensional reduction
of the $AdS_4\times S^7/Z_k$ solution of $D=11$ supergravity
described by the supercoset $OSp(8|4)/SO(7)\times SO(1,3)\times Z_k$
with 32 fermionic coordinates. The construction of
\cite{Gomis:2008jt} has generalized to superspace the results of
\cite{Giani:1984wc,Nilsson:1984bj,Sorokin:1985ap} on the relation of
$AdS_4 \times M^6$ solutions of $D=10$ type IIA supergravity and
$AdS_4\times M^7$ solutions of $D=11$ supergravity by identifying the
compact manifolds $M^7$ as $S^1$ Hopf fibrations over corresponding
$M^6$.

In  \cite{Gomis:2008jt} it has been shown that the supercoset space
$OSp(6|4)/U(3)\times SO(1,3)$ with 24 fermionic directions, which
has been used in
\cite{Arutyunov:2008if}--\cite{D'Auria:2008cw} to construct a superstring
sigma model in $AdS_4 \times CP^3$, is a subspace of the complete
superspace and that the supercoset sigma--model action (being a
partially gauge--fixed Green--Schwarz superstring action) describes
only a subsector of the complete type IIA superstring theory in
$AdS_4 \times CP^3$. The reason for this is that the
kappa--symmetry gauge fixing condition which puts to zero eight
fermionic modes corresponding to the 8 broken supersymmetries is not
admissible for all possible string configurations. So, in
particular, though the $OSp(6|4)/U(3)\times SO(1,3)$ sigma model
sector of the theory is classically integrable
\cite{Arutyunov:2008if,Stefanski:2008ik} and there are
generic arguments in favor of the integrability of the whole theory,
the direct proof of the integrability of the complete $AdS_4 \times
CP^3$ superstring still remains an open problem.

The knowledge of the explicit structure of the $AdS_4 \times CP^3$
superspace with 32 fermionic directions allows one to approach this
and other problems. The form of the string action in the $AdS_4
\times CP^3$ superspace can be drastically simplified by choosing a
suitable description of the background supergeometry and an
appropriate kappa--symmetry gauge, as was shown previously for the
cases of the type IIB superstring, D3, M2 and M5--branes in the
corresponding $AdS\times S$ backgrounds
\cite{Kallosh:1998qv}--\cite{Pasti:1998tc}. A superconformal
realization and a kappa--symmetry gauge fixing of the $OSp(6|4)$
sigma model sector of the $AdS_4 \times CP^3$ superstring have been
considered in \cite{Uvarov:2008yi} and in a light--cone gauge in
\cite{Zarembo:2009au}.

 In this paper we perform an alternative
$\kappa$--symmetry gauge fixing of the complete $AdS_4\times CP^3$
superspace which is suitable for studying regions of the theory that
are not reachable by the supercoset sigma model. In Subsection
\ref{FS} we apply this gauge fixing to simplify the superstring
action in $AdS_4\times CP^3$ and consider its T--dualization along a
$3d$ translationally invariant subspace of $AdS_4$, similar to that
performed in
\cite{Kallosh:1998ji},
which results in a simple action that contains fermions only up to
the fourth order. We also argue that, in contrast to the
$AdS_5\times S^5$ superstring
\cite{Ricci:2007eq,Berkovits:2008ic,Beisert:2008iq}, it is not possible to T--dualize
the fermionic sector of the superstring action in $AdS_4\times
CP^3$, which agrees with the conclusion of \cite{Adam:2009kt}
regarding the $OSp(6|4)$ supercoset subsector of the theory.

In addition to the superstring, also for certain configurations of
type IIA branes, \emph{e.g.} D0-- and D2--branes considered in
Section 4, the complete $AdS_4 \times CP^3$ superspace should be
used. An interesting example is a 1/2 BPS probe D2--brane placed at
the $d=3$ Minkowski boundary of $AdS_4$. Upon gauge fixing
worldvolume diffeomorphisms and kappa--symmetry, the effective
theory on the worldvolume of this D2--brane, which describes its
fluctuations in $AdS_4
\times CP^3$, is an interacting $d=3$ gauge Born--Infeld--matter theory
possessing the (spontaneously broken) superconformal symmetry
$OSp(6|4)$. The model is superconformally invariant in spite of the
presence on the $d=3$ worldvolume of the dynamical Abelian vector
field, since the latter is coupled to the $3d$ dilaton field
associated with the radial direction of $AdS_4$. The superconformal
invariance is spontaneously broken by a non--zero expectation value
of the dilaton. This example is a type IIA counterpart of so called
singleton M2, tripleton M5 and doubleton D3--branes
\cite{deWit:1998tk,Claus:1998fh,Metsaev:1998hf,Pasti:1998tc} at the boundary of $AdS_{p+2}\times S^{D-p-2}$ ($p=2,3$ and
5), respectively, in  $D=11$ supergravity and type IIB string theory
(see
\cite{Duff:2008pa} for a corresponding brane scan and a review of
related earlier work).

Another example of interest for the study of the $AdS_4/CFT_3$
correspondence is a D2--brane filling $AdS_2\times S^1\subset AdS_4$
\cite{Drukker:2008jm}. This BPS D2--brane configuration corresponds
to a disorder loop operator in the ABJM theory. Other D--brane
configurations, which are to be related to Wilson loop operators in
the ABJM theory, were considered \emph{e.g.} in
\cite{Berenstein:2008dc}--\cite{Rey:2008bh}.
In this paper we extend the bosonic action for a D2--brane wrapping
$AdS_2\times S^1$ to include the worldvolume fermionic modes.

We start our consideration with an overview of the geometry of the
$AdS_4\times CP^3$ superspace.

\section{$AdS_4 \times CP^3$ superspace}\label{superspace}

The superspace under consideration contains $AdS_4\times CP^3$
as its bosonic subspace and has 32 fermionic directions
\cite{Gomis:2008jt}. It is parametrized by the supercoordinates
\be\label{Z}
Z^{\mathcal M}=(x^{\hat
m},y^{m'},\Theta^{\underline\alpha})=(x^{\hat
m},y^{m'},\vartheta^{\alpha a'},\upsilon^{\alpha i}),
\ee
where $x^{\hat m}$ $(\hat m=0,1,2,3)$ and $y^{m'}$ $(m'=1,\cdots,6)$
are, respectively, the coordinates of $AdS_4=SO(2,3)/SO(1,3)$ and
$CP^3=SU(4)/SU(3)\times U(1)$. $\Theta^{\underline\alpha}$ are the
32 fermionic coordinates which we split into the 24 coordinates
$\vartheta^{\alpha a'}$, that correspond to the 24 unbroken
supersymmetries in the $AdS_4\times CP^3$ background, and the 8
coordinates $\upsilon^{\alpha i}$ which correspond to the broken
supersymmetries. The indices $\alpha=1,2,3,4$ are $AdS_4$ spinor
indices, $a'=1,\cdots,6$ correspond to a six--dimensional
representation of $SU(3)$ (note that the index $a'$ appearing on
spinors is different from the same index appearing in bosonic
quantities, see Appendix A.5) and $i=1,2$ are $SO(2)\sim
U(1)$ indices. For more details of our notation and conventions see
Appendix A
\footnote{Our notation and conventions are close to those in
\cite{Gomis:2008jt}. The difference is that, in this paper we put a
``hat" on the $AdS_4$ vector indices and use a more conventional IIA
superspace torsion constraint $T_{\underline{\alpha\beta}}{}^A
=-2i\Gamma^A_{\underline{\alpha\beta}}$ (instead of
$T_{\underline{\alpha\beta}}{}^A
=2\Gamma^A_{\underline{\alpha\beta}}$) and corresponding constraints
on the gauge field strengths. We also restore the dependence of the
geometric quantities and fields on the $S^7$ radius $R$, the
eleven-dimensional Planck length
$l_p=e^{\frac{1}{3}<\phi>}\sqrt{\alpha'}$ and the Chern--Simons
level $k$, which were put equal to one in
\cite{Gomis:2008jt}.}. For
the reader's convenience, below we list some of the notation used in
the text:
\begin{enumerate}
\item $D=10$ $AdS_4\times CP^3$ superspace with 24 fermions is the
supercoset $OSp(6|4)/ U(3)\times SO(1,3)$. The supervielbeins and
connections are denoted by
\begin{equation}\label{notaA}
\Big(E^{\hat a}, E^{a'}, E^{\alpha a'}, \Omega^{\hat a\hat b}, \Omega^{a'b'}, A \Big)
\end{equation}
whose expressions are given in Appendix B, eq. (B.1).
\item $D=11$ $AdS_4\times S^7$ superspace with 24 fermions. This is obtained as a $U(1)$ bundle over
the $OSp(6|4)/ U(3)\times SO(1,3)$ supercoset with the fiber
coordinate denoted by $z$. It is the supercoset $OSp(6|4)\times U(1)
/ U(3) \times SO(1,3)$ whose supervielbeins and connections are
denoted by
\begin{equation}\label{notaB}
\Big(\hat E^{\hat a}, \hat E^{a'}, \hat E^7, \hat E^{\alpha a'}, \hat \Omega^{\hat a\hat b}, \hat
\Omega^{a'b'}\Big)\,.
\end{equation}
They are given in eqs. (\ref{24thA}), see also
\cite{Gomis:2008jt}. ${\hat E}^7$ stands for the 7th (fiber) direction of $S^7$
(or, equivalently, the 11th direction in $D=11$).
\item $D=11$ $AdS_4\times S^7$ superspace  with 32 fermions.
This is the supercoset $OSp(8|4)/SO(7)\times SO(1,3)$. Its
supervielbeins and connections are denoted by
\begin{equation}\label{notaC}
\Big(\underline E^{\hat a}, \underline E^{a'}, \underline E^7, \underline E^{\alpha a'},\underline E^{\alpha i},
\underline \Omega^{\hat a\hat b},
\underline \Omega^{a'b'}, \underline \Omega^{a' 7}\Big)\,.
\end{equation}
Their explicit expressions are given in (\ref{upsilonfunctions}), (\ref{ads4connection}) and (\ref{so7connection}).
\item Finally, the $D=10$ $AdS_4\times CP^3$ superspace
with 32 fermionic directions is obtained by performing a rotation of
(\ref{notaC}) in the $(\hat a, 7)$--plane accompanied by the
dimensional reduction to $D=10$ (see \cite{Gomis:2008jt}). The
geometric quantities characterizing this superspace are denoted by
\begin{equation}\label{notaD}
\Big({\cal E}^{\hat a}, {\cal E}^{a'}, {\cal E}^{\alpha a'}, {\cal E}^{\alpha i},
{\mathcal O}^{\hat a\hat b}, {\mathcal O}^{a'b'}, {\cal A} \Big).
\end{equation}
The supervielbeins have the following form
\end{enumerate}
\be\label{simplA}
\begin{aligned}
{\mathcal E}^{a'}(x,y,\vartheta,\upsilon)&=e^{\frac{1}{3}\phi(\upsilon)}\,\left(E^{a'}(x,y,\vartheta)+2i\upsilon\,{{\sinh m}\over
m}\gamma^{a'}\gamma^5\,E(x,y,\vartheta)\right) \,,
\\
\\
{\mathcal E}^{\hat a}(x,y,\vartheta,\upsilon) &=
e^{{1\over3}\phi(\upsilon)}\,\left(E^{\hat
b}(x,y,\vartheta)+4i\upsilon\gamma^{\hat b}\,{{\sinh^2{{\mathcal M}/
2}}\over{\mathcal M}^2}\,D\upsilon\right)\Lambda_{\hat b}{}^{\hat
a}(\upsilon)
\\
&{}
\hskip+1cm -e^{-{1\over3}\phi(\upsilon)}\,\frac{R^2}{kl_p}\left(A(x,y,\vartheta)-\frac{4}{R}\upsilon\,\ve\gamma^5\,{{\sinh^2{{\mathcal
M}/2}}\over{\mathcal M}^2}\,D\upsilon\right) E_7{}^{\hat
a}(\upsilon)\,,
\\
\\
{\mathcal E}^{\alpha i}(x,y,\vartheta,\upsilon) &=
e^{{1\over6}\phi(\upsilon)}\,\left({{\sinh{\mathcal
M}}\over{\mathcal M}}\,D\upsilon\right)^{\beta j}\,S_{\beta
j}{}^{\alpha i}\,(\upsilon)
-ie^{\phi(\upsilon)}{\mathcal A}_1(x,y,\vartheta,\upsilon)\,(\gamma^5\varepsilon\lambda(\upsilon))^{\alpha
i}\,,
\\
\\
{\mathcal E}^{\alpha a'}(x,y,\vartheta,\upsilon) &=
e^{{1\over6}\phi(\upsilon)}\,E^{\gamma b'}(x,y,\vartheta)\,\left(
\delta_{\gamma}{}^{\beta}-\frac{8}{R}\,\left(\gamma^5\,\upsilon\,{{\sinh^2{{m}/2}}\over{m}^2}\right)_{\gamma
i}\upsilon^{\beta i} \right)S_{\beta b'}{}^{\alpha
a'}\,(\upsilon)\,.
\end{aligned}
\ee
The new objects appearing in these expressions, $m$, $\mathcal M$,
$\Lambda_{\hat a}{}^{\hat b}$, $E_7{}^{\hat a}$ and
$S_{\underline\alpha}^{\underline\beta}$, are functions of $\ups$
and their explicit forms are given in Appendix B.1 while the dilaton
$\phi$, dilatino $\lambda$ and RR one--form $\mathcal A_1$ are given
below. Contracted spinor indices have been suppressed, \emph{e.g.}
$(\ups\ve\gamma^5)_{\alpha i}=\ups^{\beta
j}\ve_{ji}\gamma^5_{\beta\alpha}$, where
$\varepsilon_{ij}=-\varepsilon_{ji}$, $\varepsilon_{12}=1$ is the
$SO(2)$ invariant tensor. The covariant derivative is defined as
\bee\label{D}
D\upsilon=\left(d+\frac{i}{R}E^{\hat
a}(x,y,\vartheta)\,\gamma^5\gamma_{\hat a}-\frac{1}{4}\Omega^{\hat a
\hat b}(x,y,\vartheta)\,\gamma_{\hat a\hat b}\right)\upsilon \,.
\eee
The type IIA RR one--form gauge superfield is
\be\label{simplB}
\begin{aligned}
{\mathcal A}_1(x,y,\vartheta,\upsilon) &=
R\,e^{-{4\over3}\phi(\upsilon)}\,\left[
\left(A(x,y,\vartheta)-\frac{4}{R}\upsilon\,\ve\gamma^5\,{{\sinh^2{{\mathcal
M}/2}}\over{\mathcal
M}^2}\,D\upsilon\right)\frac{R}{kl_p}\,\Phi(\upsilon)
\right.\\
&\left.\hspace{40pt}+\frac{1}{kl_p}\left(E^{\hat
a}(x,y,\vartheta)+4i\upsilon\gamma^{\hat a}\,{{\sinh^2{{\mathcal
M}/2}}\over{\mathcal M}^2}\,D\upsilon\right)E_{7\hat a}(\upsilon)
\right]\,.
\end{aligned}
\ee

The RR four-form and the NS--NS three-form superfield strengths are
given by
\be\label{f4h3}
\begin{aligned}
F_4&=d{\mathcal A}_3-{\mathcal A}_1\,H_3=-\frac{1}{4!}{\mathcal
E}^{\hat d}{\mathcal E}^{\hat c}{\mathcal E}^{\hat b}{\mathcal
E}^{\hat a}\left(\frac{6}{kl_p}\,e^{-2\phi}\Phi\ve_{\hat
a\hat b\hat c\hat d}\right) -\frac{i}{2}{\mathcal E}^{B}{\mathcal
E}^{A}{\mathcal E}^{\underline\beta}
{\mathcal E}^{\underline\alpha}e^{-\phi}(\Gamma_{AB})_{\underline{\alpha\beta}}\,,\\
H_3&=dB_2=-\frac{1}{3!}{\mathcal E}^{\hat c}{\mathcal E}^{\hat
b}{\mathcal E}^{\hat a}(\frac{6}{kl_p}e^{-\phi}\ve_{\hat a\hat b\hat
c\hat d}E_7{}^{\hat d}) -i{\mathcal E}^{A}{\mathcal
E}^{\underline\beta}{\mathcal
E}^{\underline\alpha}(\Gamma_A\Gamma_{11})_{\underline{\alpha\beta}}
+i{\mathcal E}^{B}{\mathcal E}^{A}{\mathcal
E}^{\underline\alpha}(\Gamma_{AB}\Gamma^{11}\lambda)_{\underline\alpha}
\end{aligned}
\ee
and the corresponding gauge potentials are
\be\label{B2}
B_2=b_2+\int_0^1\,dt\,i_\Theta H_3(x,y,t\Theta)\,,\qquad \Theta=(\vartheta,\upsilon)\,\\
\ee
\be\label{A3}
\hskip+1.9cm{\mathcal
A}_3=a_3+\int_0^1\,dt\,i_\Theta\left(F_4+\mathcal{A}_1H_3\right)(x,y,t\Theta)\,,
\ee
where $b_2$ and $a_3$ are the purely bosonic parts of the gauge
potentials and $i_\Theta$ means the inner product with
$\Theta^{\underline\alpha}$. Note that $b_2$ is  pure gauge in the
$AdS_4\times CP^3$ solution while $a_3$ is the RR three-form
potential of the bosonic background.

 The dilaton superfield $\phi(\upsilon)$, which depends only on
the eight fermionic coordinates corresponding to the broken
supersymmetries, has the following form in terms of $E_7{}^{\hat
a}(\upsilon)$ and $\Phi(\upsilon)$
\be\label{dilaton1}
e^{{2\over
3}\phi(\upsilon)}={R\over{kl_p}}\,\sqrt{\Phi^2+E_7{}^{\hat
a}\,E_7{}^{\hat b}\,\eta_{\hat a\hat b}}\,.
\ee
The value of the dilaton at $\upsilon=0$ is
\begin{equation}
e^{\frac{2}{3}\phi(\upsilon)}|_{\upsilon=0}=e^{\frac{2}{3}\phi_0}=\frac{R}{kl_p}\,.
\end{equation}
The fermionic field $\lambda^{\alpha i}(\upsilon)$ describes the
non--zero components of the dilatino superfield and is given by the
equation \cite{Howe:2004ib}
\be\label{dilatino1}
\lambda_{\alpha i}=-\frac{i}{3}D_{\alpha i}\,\phi(\upsilon)\,.
\ee

In the above expressions $E^{\hat a}(x,y,\vartheta)$,  $E^{
a'}(x,y,\vartheta)$ and $\Omega^{\hat a\hat b}(x,y,\vartheta)$ are
the supervielbeins and the $AdS_4$ part of the spin connection of
the supercoset $OSp(6|4)/U(3)\times SO(1,3)$ and $A(x,y,\vartheta)$
is the corresponding type IIA RR one--form gauge superfield, eq.
(\ref{notaA}), whose explicit form is given in Appendix B.

As mentioned above other quantities appearing in eqs.
(\ref{simplA})--(\ref{dilatino1}), namely $\mathcal M$, $m$,
$\Phi(\ups)$, $E_7{}^{\hat a}(\ups)$, $\Lambda_{\hat a}{}^{\hat b}(\ups)$ and
$S_{\underline\beta}{}^{\underline\alpha}(\ups)$, whose geometrical and
group--theoretical meaning has been explained in
\cite{Gomis:2008jt}, are also given in Appendix B.

\setcounter{equation}0
\section{Kappa--symmetry gauge fixing}
We shall now consider conditions for gauge fixing kappa--symmetry
which are convenient for the description of configurations of
superstrings and D-branes in the $AdS_4\times CP^3$ superbackground
described above and for studying $AdS_4/CFT_3$ correspondence
problems.

Since the $AdS_4/CFT_3$ holography is realized at the $3d$ Minkowski
boundary of $AdS_4$ it is convenient to choose the $AdS_4\times
CP^3$ metric in the form
\be\label{ads4metric11}
ds^2=\left(r\over
{R_{{CP^3}}}\right)^4\,dx^m\,\eta_{mn}\,dx^n+\left({R_{CP^3}}\over
r\right)^2\,dr^2+R_{CP^3}^2\,ds^2_{_{CP^3}}\,
\ee
where $m=0,1,2$ are indices corresponding to the coordinates of the
$3d$ Minkowski boundary and $r$ is the 4th, radial, coordinate of
$AdS_4$. So the $AdS_4$ coordinates are $x^{\hat m}=(x^m,r)$. The
$AdS_4$ radius is half of the $CP^3$ radius $R_{CP^3}$ which (in the
string frame) is related to the $S^7$ radius $R$ as follows
\be\label{R}
R_{CP^3} = e^{\frac{1}{3}\phi_0}R =\left(\frac{R^3}{kl_p}\right)^{1/2}\,.
\ee

In the coordinate system associated with the metric
(\ref{ads4metric11}) (the bosonic part of) the RR field ${\mathcal A}_3$, whose flux,
together with $F_2=da_1={e^{-\phi_0}\over
{R_{CP^3}}}\,dy^{m'}dy^{n'}J_{m'n'}$ (where $dy^{m'}dy^{n'}J_{m'n'}$
is the K\"ahler form on $CP^3$), ensures the compactification on
$AdS_4\times CP^{3}$
\cite{Watamura:1983hj,Nilsson:1984bj,Sorokin:1985ap}, has the
following form
\be\label{A31}
a_3=e^{-\phi_0}\left({r\over
{R_{CP^3}}}\right)^6\,dx^0\,dx^1\,dx^2\,,\qquad F_4={6\over
R_{CP^3}}e^{-\phi_0}\,\left({r\over
{R_{CP^3}}}\right)^5\,dx^0\,dx^1\,dx^2\,dr\,.
\ee
(In our conventions the exterior derivative acts from the right.)

Instead of the $AdS_4$ part of the metric (\ref{ads4metric11}),
which obscures a bit the fact that the metric of the conformal
boundary is the flat Minkowski metric on $R^{1,2}$, one can use the
$AdS_4$ metric in the conformally flat form
\be\label{ads4metric21}
ds^2_{_{AdS_4}}={1\over
u^2}(dx^m\eta_{mn}dx^n+\frac{R_{CP^3}^2}{4}\,du^2)\,,
\qquad u=\left(R_{CP^3}\over r\right)^2\,.
\ee
This metric is associated with a simple coset representative $
g=\exp(x^m\,\Pi_m)$ $\exp(R_{CP^3}\,\ln(u) D)$, where $\Pi_m$ are
the generators of the Poincar\'e translations along the Minkowski
boundary $([\Pi_m,\,\Pi_n]=0)$ and $D$ is the dilatation generator
$[D,\,\Pi_m]=\Pi_m$.

Note that if the components of the vielbein associated with the
metric (\ref{ads4metric11}) or (\ref{ads4metric21}) are chosen to
be\footnote{Note that the vielbeins $e^a$ and $e^3$ appearing in eq.
(\ref{ad4v}) correspond to the $AdS_4$ metric of the $D=11$
$AdS_4\times S^7$ solution characterized by the radius R which is
related to the $CP^3$ radius in the string frame according to eq.
(\ref{R}). These bosonic vielbeins will appear in our explicit
expressions for the $AdS_4\times CP^3$ supergeometry.}
\be\label{ad4v}
e^{\frac{\phi_0}{3}}\,e^a={r^2\over
R_{CP^3}^2}\,dx^a=u^{-1}\,dx^a\,,\qquad
e^{\frac{\phi_0}{3}}\,e^3=\frac{R_{CP^3}}{r}\,dr=-\frac{R_{CP^3}}{2u}\,du,
\ee
the components of the $SO(1,3)$ spin connection are
\be\label{eaoa31}
\omega^{a3}=-\frac{2}{R}\,e^a\,,
\ee
and
\be\label{oab}
\omega^{ab}=0\,.
\ee
We shall use the relation (\ref{eaoa31}) to simplify the form of the
gauge fixed $AdS_4 \times CP^3$ supergeometry. Note that the
condition (\ref{eaoa31}) can always be imposed by performing an
appropriate local $SO(1,3)$ transformations of the vielbein and
connection, though in general the $SO(1,2)$ components $\omega^{ab}$
of the connection will be non--zero.

Using the previous experience of gauge fixing kappa--symmetry of superstrings, D-branes and M-branes in AdS backgrounds
\cite{Kallosh:1998qv}--\cite{Pasti:1998tc} we choose the
kappa--symmetry gauge fixing condition in the form
\footnote{Such a gauge for fixing kappa--symmetry is analogous to
the so called Killing spinor gauge \cite{Kallosh:1998qv}, or
supersolvable gauge
\cite{Dall'Agata:1998wz}, or the superconformal gauge \cite{Pasti:1998tc}.}
\be\label{kappagauge1}
\Theta=\frac{1}{2}(1\pm\gamma)\Theta\,\quad \Rightarrow
\quad \vartheta^{a'}=\frac{1}{2}(1\pm\gamma)\vartheta^{a'}\,,\qquad
\upsilon^{i}=\frac{1}{2}(1\pm\gamma)\upsilon^{i},
\ee
where
\be\label{gamma}
\gamma=\gamma^{012}\qquad\Rightarrow\qquad\gamma^2=1,
\quad\{\gamma,\gamma^5\}=[\gamma,\gamma^a]=0\quad\mbox{and}\quad\gamma\gamma^3=-i\gamma^5\,,
\ee
$a=0,1,2$ are the indices of the 3d Minkowski boundary or of
$AdS_2\times S^1$ and $\gamma^3$ is associated with the third
spatial direction of $AdS_4$. Note that, in view of our definition
(\ref{Gamma10}) of the $D=10$ gamma--matrices the matrices defined
in (\ref{gamma}) can be regarded either as $4d$ gamma matrices or as
the $D=10$ matrices $\Gamma^{\hat a}=\gamma^{\hat a}\otimes {\bf 1}$
$(\hat a=0,1,2,3)$.

The condition (\ref{kappagauge1}) is admissible for fixing
kappa--symmetry if the projection matrix $\frac{1}{2}(1\mp\gamma)$ either coincides or does not
commute with the kappa--symmetry projection matrix $\frac{1}{2}(1+\Gamma)$ of a given
configuration of the superstring and D--branes.
This can be understood in the following way. To lowest order in fermions $\Theta$ transforms under kappa--symmetry as
\be
\delta_\kappa\Theta=\frac{1}{2}(1+\Gamma)\kappa\,,
\ee
where $\frac{1}{2}(1+\Gamma)$ is a projection matrix and
$\kappa(\xi)$ is an arbitrary spinor parameter. It is then clear
that if the two projectors coincide, we can pick a $\kappa$ such
that $\frac{1}{2}(1+\Gamma)\Theta=0$, or equivalently
$\Theta=\frac{1}{2}(1-\Gamma)\Theta$. In the case when the two
projection operators do not coincide a kappa--symmetry variation of
the gauge--fixing condition $\frac{1}{2}(1\mp\gamma)\Theta=0$ which
leaves it intact gives
\be
0=\frac{1}{4}(1\mp\gamma)(1+\Gamma)\kappa
=\frac{1}{8}(1+\Gamma)(1\mp\gamma)(1+\Gamma)\kappa\mp\frac{1}{8}[\gamma,\Gamma](1+\Gamma)\kappa
=\mp\frac{1}{8}[\gamma,\Gamma](1+\Gamma)\kappa\,,
\ee
where in the last step we made use of the initial equation. This
means that for the gauge--fixing to be complete, i.e. that the
variation of the gauge fixing condition vanishes if and only if all
independent kappa--symmetry parameters are put to zero, the
commutator $[\gamma,\Gamma]$ has to be an invertible matrix (when
restricted to the relevant subspace).

 As we shall see
below, for any choice of the sign the condition (\ref{kappagauge1})
    is an admissible gauge-fixing in the case of
    arbitrary motion of D0--branes in $AdS_4 \times CP^3$, while in the
case of the superstring it is admissible (for both signs) for those
 configurations for which the projection of the string
worldsheet on the $3d$ Minkowski boundary is a non--degenerate
two--dimensional time--like surface. In the case of the D2--brane
placed at the Minkowski boundary of $AdS_4$, to gauge fix kappa--symmetry one must choose the condition (\ref{kappagauge1}) with the
lower sign
\cite{Pasti:1998tc}, while both signs are admissible when the
    D2--brane wraps an $AdS_2 \times S^1$ subspace of $AdS_4$. However,
    the choice of (\ref{kappagauge1}) with the upper sign yields
    the simplest gauge-fixed form of the string and
    brane actions in the $AdS_4\times CP^3$ superbackground.

When the fermionic coordinates are restricted by the condition
(\ref{kappagauge1}), the expressions for the supervielbeins and the
gauge superfields of the $AdS_4 \times CP^3$ superspace drastically
simplify due to the identities satisfied by the projected fermionic
coordinates given in Appendix C. In particular, the functions of
$\upsilon$ which enter the eqs. (\ref{simplA})--(\ref{dilaton1}),
whose explicit forms are given in Appendix B.1, reduce to
\be\label{Phi1}
\Phi(\upsilon)=1+\frac{8}{R}\,\upsilon\,\ve\gamma^5\,{{\sinh^2{{\mathcal M}/2}}\over{\mathcal
M}^2}\,\ve\upsilon =1\,,
\ee
\bee
E_7{}^a(\upsilon)&=&-\frac{8i}{R}\,\upsilon\gamma^a\,{{\sinh^2{{\mathcal M}/2}}\over{\mathcal M}^2}\,\varepsilon\,{\upsilon}=-\frac{2i}{R}\ups\gamma^a\ve\ups\,,\\
E_7{}^3(\upsilon)&=&-\frac{8i}{R}\,\upsilon\gamma^3\,{{\sinh^2{{\mathcal
M}/2}}\over{\mathcal M}^2}\,\varepsilon\,{\upsilon}=0\,.
\eee
The dilaton superfield (\ref{dilaton1}) takes the form
\be\label{gfdialton}
e^{{2\over 3}\phi(\upsilon)}
=\frac{R}{kl_p}(1-\frac{6}{R^2}(\ups\ups)^2)
\quad \Rightarrow \quad
\phi(\upsilon)=\frac{3}{2}(\log\frac{R}{kl_p}-\frac{6}{R^2}(\ups\ups)^2)\,,
\ee
where $\ups\ups=\delta_{ij}C_{\alpha\beta}\ups^{\alpha i}\ups^{\beta j}$, and the dilatino becomes
\bee
\lambda^{\alpha i}
=
\frac{2i}{R}\left(\frac{R}{kl_p}\right)^{-1/4}((\gamma^5\upsilon)^{\alpha i}+\frac{3}{R}\ups^{\alpha i}\,\ups\ups)\,.
\eee
We also find that
\be
\Lambda_a{}^b
= (1+\frac{2}{R^2}(\ups\ups)^2)\delta_a{}^b\,,
\qquad
\Lambda_3{}^a = \Lambda_a{}^3=0\,,\qquad
\Lambda_3{}^3=1\,,
\ee
and
\bee\label{S}
&&S_{\underline\alpha}{}^{\underline\beta}=
\left(\frac{R}{kl_p}\right)^{1/4}e^{-\frac{1}{6}\phi}\,\delta_{\underline\alpha}{}^{\underline\beta}
+\frac{i}{R}\ups\gamma^a\ve\ups\,(\Gamma_a\Gamma_{11})_{\underline\alpha}{}^{\underline\beta}\,.
\eee

\subsection{$AdS_4 \times CP^3$ supergeometry with
$\Theta={1\over 2}(1+\gamma)\Theta$}\label{theta-}

The supervielbeins (\ref{simplA}) and the gauge superfields
(\ref{simplB}), (\ref{B2}) and (\ref{A3}) take the simplest form
when the kappa--symmetry gauge condition (\ref{kappagauge1}) is
chosen with the upper sign. In virtue of eqs.
(\ref{Phi1})--(\ref{S}) and expressions given in Appendix C, the
supervielbeins reduce to
\be\label{simple+v}
\begin{aligned}
{\mathcal
E}^{a'}(x,y,\vartheta,\upsilon)&=\Big(\frac{R}{kl_p}\Big)^{1/2}e^{a'}(y)(1-\frac{3}{R^2}(\ups\ups)^2)\,,
\\
\\
{\mathcal E}^a(x,y,\vartheta,\upsilon) &=\Big(\frac{R}{kl_p}\Big)^{1/2}
(e^a(x)+i\Theta\gamma^aD\Theta)(1-\frac{1}{R^2}(\ups\ups)^2)\,,
\\
\\
{\mathcal E}^3(x,y,\vartheta,\upsilon)
&=\Big(\frac{R}{kl_p}\Big)^{1/2}e^3(x)(1-\frac{3}{R^2}(\ups\ups)^2)\,,
\end{aligned}
\ee
\be\label{simplesv}
\begin{aligned}
{\mathcal E}^{\alpha i}(x,y,\vartheta,\upsilon) &=
\Big(\frac{R}{kl_p}\Big)^{1/4}\Big( (D_8\ups)^{\alpha
i}-\frac{1}{R}\ups\gamma^a\ve\ups\,(D_8\ups\ve\gamma_a\gamma_5)^{\alpha
i}
-\frac{4i}{R^2}(e^a(x)+i\Theta\gamma^aD\Theta)(\gamma_a\ups)^{\alpha
i}\,\ups\ups
\Big)
\,,
\nonumber\\
\nonumber\\
{\mathcal E}^{\alpha a'}(x,y,\vartheta,\upsilon) &=
\Big(\frac{R}{kl_p}\Big)^{1/4}\Big( (D_{_{24}}\vartheta)^{\alpha a'}
+\frac{i}{R}\ups\gamma^a\ve\ups\,(D_{_{24}}\vartheta\gamma_a\gamma_5\gamma_7)^{\alpha
a'}\Big)\,.\nonumber
\end{aligned}
\ee
The type IIA RR one--form gauge superfield is
\be\label{simple+A}
\begin{aligned}
{\mathcal A}_1(x,y,\vartheta,\upsilon) &=kl_p\Big(
A(y)-\frac{2i}{R^2}(e^a(x)+i\Theta\gamma^aD\Theta)\,\ups\ve\gamma_a\ups\Big)\,,
\end{aligned}
\ee
where $A(y)$ is the potential for the K\"ahler form on $CP^3$,
\emph{i.e.} $dA(y)=\frac{1}{R^2}\,dy^{m'}dy^{n'}\,J_{m'n'}$, and the covariant derivatives are
\bee\label{D81}
D\Theta&=&(D_8\upsilon,\,D_{24}\vartheta)\\
D_8\upsilon&=&{\mathcal P_2}\,\left(d-\frac{1}{R}e^3-\frac{1}{4}\omega^{ab}\gamma_{ab}+2A(y)\ve\right)\upsilon\nonumber\\
D_{24}\vartheta&=&{\mathcal P}_6\,(d-\frac{1}{R}e^3
+\frac{i}{R}e^{a'}\gamma_{a'} -\frac{1}{4}\omega^{ab}\gamma_{ab}
-\frac{1}{4}\omega^{a'b'}\gamma_{a'b'})\vartheta\,,\nonumber
\eee
where ${\mathcal P}_2$ and ${\mathcal P}_6$ are projectors that
single out from 32 $\Theta$, respectively, 8 $\upsilon$ and 24
$\vartheta$ (see Appendix A.5). The appearance of the $U(1)$ gauge potential $A(y)$ in the covariant
derivative of $\upsilon$ (\ref{D81}) reflects the fact that
$\upsilon$ has $U(1)$ charge equal to 2.
\\
Note that
\be
D_8={\mathcal P}_2\, {\mathcal D}\,{\mathcal P}_2\,,\qquad
D_{24}={\mathcal P}_6\, {\mathcal D}\,{\mathcal P}_6\,,
\ee
where
\be
{\mathcal D}=d-\frac{1}{R}e^3+\frac{i}{R}e^{a'}\gamma_{a'}
-\frac{1}{4}\omega^{ab}\gamma_{ab}-\frac{1}{4}\omega^{a'b'}\gamma_{a'b'}\,.
\ee
The NS--NS three-form, eq. (\ref{f4h3}), becomes
\bee\label{H3ex}
H_3&=&-\frac{6i}{R^2} e^3\, {\mathcal E}^b\,{\mathcal
E}^a\,\varepsilon_{abc}\,\upsilon\gamma^c\varepsilon\upsilon
-\frac{2R}{kl_p}\,\Big[
\frac{i}{R}(e^b+i\Theta\gamma^bD\Theta)(e^a+i\Theta\gamma^aD\Theta)\,D_8\upsilon\gamma_{ab}\varepsilon\upsilon
\nonumber\\
&&{}
+\frac{1}{R}(e^a+i\Theta\gamma^aD\Theta)\,D\Theta\,\gamma_{ab}\,D\Theta\,\ups\gamma^b\varepsilon\ups
+\frac{1}{2}\,e^3\,D\Theta\,\gamma_7\,D\Theta
+\frac{2i}{R}\,e^3\,e^{a'}\,D\Theta\gamma_{a'}\gamma_7\upsilon
\nonumber\\
&&{} +\frac{i}{2}\,e^{a'}\,D\Theta\,\gamma_{a'}\gamma_7\,D\Theta
+\frac{1}{R}\,e^{b'}\,e^{a'}\,D\Theta\,\gamma_{a'b'}\gamma_7\upsilon\,\Big]\,,
\eee
where $\Theta=(\vartheta,\upsilon)$ and
$D\Theta=(D_{24}\vartheta,\,D_8\upsilon)$.
\\
We now want to determine the potential of $H_3=dB_2$ using eq.
(\ref{B2}). Taking into account that $i_\Theta\mathcal E^{A}=0$, and
the fact that with the plus sign in the projector
(\ref{kappagauge1})
$\Theta\gamma_{ab}D\Theta=\ve_{abc}\,\Theta\gamma^cD\Theta$ etc., we
get
\bee
i_\Theta H_3 &=& 2\frac{R}{kl_p}
\Big(
-\frac{i}{R}e^be^a\,\ups\gamma_{ab}\ve\ups
-\frac{2i}{R}e^3e^{a'}\,\vartheta\gamma_{a'}\gamma^7\ups
-\frac{1}{R}e^{b'}e^{a'}\,\Theta\gamma_{a'b'}\gamma^7\ups
+e^3\,\Theta\gamma^7D\Theta
\nonumber\\
&&{} +ie^{a'}\,\Theta\gamma_{a'}\gamma^7D\Theta
+\frac{4}{R}e^b\,\Theta\gamma^aD\Theta\,\ups\gamma_{ab}\ve\ups
+\frac{3i}{R}\Theta\gamma^bD\Theta\,\Theta\gamma^aD\Theta\,\ups\gamma_{ab}\ve\ups
\Big)\,.
\eee
This gives the NS--NS two-form potential (see eq. (\ref{B2}))
\bee\label{simple+B}
B_2&=&
\frac{R}{kl_p}
\Big[
-\frac{i}{R}(e^b+i\Theta\gamma^bD\Theta)\,(e^a+i\Theta\gamma^aD\Theta)\,
\ups\gamma_{ab}\ve\ups
-\frac{2i}{R}e^3e^{a'}\,\vartheta\gamma_{a'}\gamma^7\ups
\nonumber\\
&&{\hspace{20pt}}
-\frac{1}{R}e^{b'}e^{a'}\,\Theta\gamma_{a'b'}\gamma^7\ups
+e^3\,\Theta\gamma^7D\Theta
+ie^{a'}\,\Theta\gamma_{a'}\gamma^7D\Theta
\Big]\,.
\eee

Now we turn our attention to the RR four-form $F_4$ (\ref{f4h3}) and
its potential $\mathcal A_3$ (\ref{A3}). $F_4$ simplifies to
\be\label{f4}
F_4= -\frac{1}{kl_p}\,e^{-2\phi}\,{\mathcal E}^3{\mathcal
E}^c{\mathcal E}^b{\mathcal E}^a\,\ve_{abc}
-\frac{i}{2}e^{-\phi}{\mathcal E}^{B}{\mathcal E}^{A}{\mathcal
E}^{\underline\beta} {\mathcal
E}^{\underline\alpha}(\Gamma_{AB})_{\underline{\alpha\beta}}\,,
\ee
which gives
\bee
i_\Theta F_4 &=& -i(\frac{R}{kl_p})^{1/4}e^{-\phi}{\mathcal
E}^{B}{\mathcal E}^{A} ({\mathcal E}\Gamma_{AB}\Theta
+\frac{i}{R}{\mathcal
E}\Gamma_{AB}\gamma_a\Gamma_{11}\Theta\,\ups\gamma^a\ve\ups)
\nonumber\\
&=&
-i(e^b+i\Theta\gamma^bD\Theta)(e^a+i\Theta\gamma^aD\Theta)\,(1+\frac{12}{R^2}(\ups\ups)^2)(D\Theta\gamma_{ab}\Theta
+\frac{4i}{R^2}e^c\ve_{abc}\,(\ups\ups)^2 )
\nonumber\\
&&{}
+\frac{4i}{R}e^3(e^a+i\Theta\gamma^aD\Theta)\,D\Theta\gamma^7\Theta\,\ups\gamma_a\ve\ups
-\frac{4}{R}e^{a'}(e^a+i\Theta\gamma^aD\Theta)\,D\Theta\gamma_{a'}\gamma^7\Theta\,\ups\gamma_a\ve\ups
\nonumber\\
&&{} +2e^3e^{a'}(D\Theta\gamma_{a'}\Theta
+\frac{4i}{R^2}(e^a+i\Theta\gamma^aD\Theta)\ups\gamma_a\gamma_{a'}\vartheta\,\ups\ups
)
\nonumber\\
&&{} -ie^{b'}e^{a'}(D\Theta\gamma_{a'b'}\Theta
+\frac{4i}{R^2}(e^a+i\Theta\gamma^aD\Theta)\ups\gamma_a\gamma_{a'b'}\vartheta\,\ups\ups)\,.
\eee
Since $i_\Theta\mathcal A_1=0$ the RR three--form potential
(\ref{A3}) becomes
\bee\label{A3-}
\mathcal A_3&=&a_3+\int_0^1\,dt(i_\Theta F_4+\mathcal A_1i_\Theta H_3)(x,y,t\Theta)
\nonumber\\
&=& a_3 -\frac{i}{2}e^be^a\,D\Theta\gamma_{ab}\Theta
+e^3e^{a'}\,D\Theta\gamma_{a'}\Theta
-\frac{i}{2}e^{b'}e^{a'}\,D\Theta\gamma_{a'b'}\Theta
+\frac{1}{2}e^b\Theta\gamma^aD\Theta\,D\Theta\gamma_{ab}\Theta
\nonumber\\
&&{}
+\frac{i}{6}\Theta\gamma^bD\Theta\,\Theta\gamma^aD\Theta\,D\Theta\gamma_{ab}\Theta
+k\,l_p\,A(y)\,B_2\,.
\eee
Looking at the purely bosonic part of $F_4$, eq. (\ref{f4}) it is
easy to see (compare also with eqs. (\ref{A31})) that we can take
\be
a_3=\frac{1}{3!}e^ce^be^a\,\ve_{abc}\,.
\ee
Note that in the above expressions for the supervielbeins
(\ref{simple+v}), the RR one--form (\ref{simple+A}), the three--form
(\ref{A3-}) and the NS-NS two--form (\ref{simple+B}) the maximum
order of the fermions is six.

\subsection{$AdS_4 \times CP^3$ supergeometry with
$\Theta={1\over 2}(1-\gamma)\Theta$}\label{theta++}

When the condition (\ref{kappagauge1}) is chosen with the lower
sign, in view of eqs. (\ref{Phi1})--(\ref{S}) and expressions given
in Appendix C, the supervielbeins (\ref{simplA}) and the RR
one--form gauge superfield (\ref{simplB}) reduce to a form which is
more complicated than their gauge--fixed counterparts of the
previous Subsection. But, as we have already mentioned, one cannot
use the gauge fixing condition of Subsection \ref{theta-} to
describe the D2--brane at the Minkowski boundary of $AdS_4$, and
should impose $\Theta={1\over 2}(1-\gamma)\Theta$ instead. In this
case the supervielbeins take the following form
\be\label{-bv}
\begin{aligned}
{\mathcal E}^{a'}(x,y,\vartheta,\upsilon)&=\Big(\frac{R}{kl_p}\Big)^{1/2}
\Big(e^{a'}(y)-\frac{2}{R}e^a(x)\,\Theta\gamma^{a'}\gamma_a\Theta\Big)
(1-\frac{3}{R^2}(\ups\ups)^2)\,,
\\
\\
{\mathcal E}^a(x,y,\vartheta,\upsilon)
&=\Big(\frac{R}{kl_p}\Big)^{1/2}\,\Big( e^a(x) +i\Theta\gamma^a D\Theta
+\frac{1}{R^2}e^a(x)(\vartheta\vartheta-\ups\ups)^2
\Big)(1-\frac{1}{R^2}(\ups\ups)^2)\,,
\\
\\
{\mathcal E}^3(x,y,\vartheta,\upsilon)
&=\Big(\frac{R}{kl_p}\Big)^{1/2}e^3(x)(1-\frac{3}{R^2}(\ups\ups)^2)\,,
\end{aligned}
\ee
\be\label{-fv}
\begin{aligned}
{\mathcal E}^{\alpha i}(x,y,\vartheta,\upsilon) &=
\Big(\frac{R}{kl_p}\Big)^{1/4}\Big( (D_8\ups)^{\alpha i}
-\frac{1}{R}\ups\gamma^a\ve\ups\,(D_8\ups\ve\gamma_a\gamma_5)^{\alpha
i}
\\
&{}
\hskip+1cm
-\frac{4i}{R^2}(e^a(x)+i\Theta\gamma^aD\Theta
+\frac{1}{R^2}e^a(x)(\vartheta\vartheta-\ups\ups)^2
)(\gamma_a\ups)^{\alpha i}\,\ups\ups
\Big)
\,,
\\
\\
{\mathcal E}^{\alpha a'}(x,y,\vartheta,\upsilon) &=
\Big(\frac{R}{kl_p}\Big)^{1/4}\Big( (D_{_{24}}\vartheta)^{\alpha a'}
+\frac{i}{R}(D_{24}\vartheta\gamma_a\gamma^5\gamma^7)^{\alpha
a'}\,\ups\gamma^a\ve\ups
\Big)\,.
\end{aligned}
\nonumber
\ee
The type IIA RR one--form gauge superfield is
\be\label{A2-}
{\mathcal A}_1(x,y,\vartheta,\upsilon)=kl_p\Big(A(y)
-\frac{2}{R^2}e^a(x)\,\Theta\gamma^7\gamma_a\Theta
 -\frac{2i}{R^2}(e^a(x)+i\Theta\gamma^aD\Theta
+\frac{1}{R^2}e^a(x)(\vartheta\vartheta)^2
)\ups\ve\gamma_a\ups
\Big)\,.
\ee
In the above expressions
\bee\label{D8}
D\Theta&=&(D_8\upsilon\,,D_{24}\vartheta)\,,\nonumber\\
D_8\upsilon&=&(D-\frac{2i}{R^2}\ups\ups\,e^a\,\gamma_a+2A(x,y,\vartheta)\,\ve)\upsilon
\\
&&\hspace{-40pt}=
\Big(d+\frac{2i}{R}e^a(\gamma^5\gamma_a+\frac{1}{R}(\vartheta\vartheta-\ups\ups)\gamma_a)
+\frac{1}{R}e^3
-\frac{1}{4}\omega^{ab}\gamma_{ab}
+(2A(y)-\frac{4}{R^2}e^a\vartheta\gamma^7\gamma_a\vartheta)\ve
\Big)\upsilon,
\nonumber
\eee
\bee\label{D_{24}}
D_{_{24}}\vartheta
&=&{\mathcal P}_6\,\Big(d
+\frac{2i}{R}e^a(\gamma^5\gamma_a+\frac{1}{R}(\vartheta\vartheta-\ups\ups)\gamma_a)
+\frac{1}{R}e^3 +\frac{i}{R}e^{a'}\gamma_{a'}
-\frac{1}{4}\omega^{ab}\gamma_{ab}
-\frac{1}{4}\omega^{a'b'}\gamma_{a'b'}\Big)\vartheta\,.\nonumber\\
\eee
(The shift of $D$ by $-\frac{2i}{R^2}\ups\ups\,e^a\,\gamma_a$ has
been made for the expressions to have a nicer and more
covariant--looking form).
\\
The NS--NS three-form, eq. (\ref{f4h3}), becomes
\be
H_3= -\frac{6i}{R^2}e^3{\mathcal E}^b{\mathcal
E}^a\,\ve_{abc}\,\ups\gamma^c\ve\ups -i{\mathcal E}^{A}{\mathcal
E}^{\underline\beta}{\mathcal
E}^{\underline\alpha}(\Gamma_A\Gamma_{11})_{\underline{\alpha\beta}}
+i{\mathcal E}^{B}{\mathcal E}^{A}{\mathcal
E}^{\underline\alpha}(\Gamma_{AB}\Gamma^{11}\lambda)_{\underline\alpha}
\,.
\ee
We now would like to determine its potential according to eq.
(\ref{B2}). Using the fact that
\be
i_\Theta{\mathcal E}^{\underline\alpha}=
\Big(\frac{R}{kl_p}\Big)^{1/4}(\Theta^{\underline\alpha}
+\frac{i}{R}\ups\gamma^a\ve\ups\,(\Theta\Gamma_a\Gamma_{11})^{\underline\alpha})
\ee
and $i_\Theta\mathcal E^{A}=0$ we get
\bee\label{iH3-}
i_\Theta H_3&=&
\frac{R}{kl_p}\Big(
-\frac{2}{R}(e^b+i\Theta\gamma^bD\Theta+\frac{1}{R^2}e^b(\vartheta\vartheta-\ups\ups)^2)
(e^a+i\Theta\gamma^aD\Theta+\frac{1}{R^2}e^a(\vartheta\vartheta-\ups\ups)^2)\ups\gamma_{ab}\gamma^7\ups
\nonumber\\
&&{}
+\frac{4}{R}(e^a+i\Theta\gamma^aD\Theta+\frac{1}{R^2}e^a(\vartheta\vartheta-\ups\ups)^2)
D\Theta\gamma_{ba}\Theta\,\ups\gamma^b\ve\ups
+\frac{8}{R^2}e^3e^a\vartheta\vartheta\,\ups\gamma_a\ve\ups
\nonumber\\
&&{}
+\frac{4}{R}(e^a+i\Theta\gamma^aD\Theta+\frac{1}{R^2}e^a(\vartheta\vartheta-\ups\ups)^2)e^b\Theta\gamma_{ba}\gamma^7\Theta
+2e^3D\Theta\gamma^7\Theta
\nonumber\\
&&{}
-2i(e^{a'}-\frac{2}{R}e^a\,\Theta\gamma^{a'}\gamma_a\Theta)D\Theta\gamma_{a'}\gamma^7\Theta
+\frac{4i}{R}e^3(e^{a'}-\frac{2}{R}e^c\,\Theta\gamma^{a'}\gamma_c\Theta)\vartheta\gamma_{a'}\gamma^7\ups
\nonumber\\
&&{}
-\frac{2}{R}(e^{b'}-\frac{2}{R}e^b\,\Theta\gamma^{b'}\gamma_b\Theta)(e^{a'}-\frac{2}{R}e^c\,\Theta\gamma^{a'}\gamma_c\Theta)\Theta\gamma_{a'b'}\gamma^7\ups
\nonumber\\
&&{}
+\frac{8i}{R^2}(e^{a'}-\frac{2}{R}e^c\,\Theta\gamma^{a'}\gamma_c\Theta)e^a\Theta\gamma_{a'}\gamma_{ab}\Theta\,\ups\gamma^b\ve\ups
\Big)
\eee
and finally
\bee\label{B2-}
B_2&=&
\frac{R}{kl_p}\Big(
\frac{i}{R}(e^b+i\Theta\gamma^bD\Theta+\frac{1}{R^2}e^b(\vartheta\vartheta)^2)(e^a+i\Theta\gamma^aD\Theta+\frac{1}{R^2}e^a(\vartheta\vartheta)^2)\,\ve_{abc}\,\ups\gamma^c\ve\ups
\nonumber\\
&&{}
+\frac{2}{R}(e^a+\frac{i}{2}\Theta\gamma^aD\Theta+\frac{1}{3R^2}e^a(\vartheta\vartheta-\ups\ups)^2)e^b\,\ve_{abc}\,\Theta\gamma^c\gamma^7\Theta
+\frac{2i}{R}e^3(e^{a'}-\frac{1}{R}e^a\,\Theta\gamma^{a'}\gamma_a\Theta)\,\vartheta\gamma_{a'}\gamma^7\ups
\nonumber\\
&&{}
-\frac{1}{R}(e^{b'}-\frac{1}{R}e^b\,\Theta\gamma^{b'}\gamma_b\Theta)(e^{a'}-\frac{1}{R}e^a\,\Theta\gamma^{a'}\gamma_a\Theta)\,\Theta\gamma_{a'b'}\gamma^7\ups
-\frac{1}{3R^3}e^b\,\Theta\gamma^{b'}\gamma_b\Theta\,e^c\,\Theta\gamma^{a'}\gamma_c\Theta\,\Theta\gamma_{a'b'}\gamma^7\ups
\nonumber\\
&&{} -e^3\,\Theta\gamma^7D\Theta
+\frac{i}{R^2}e^3e^a(\vartheta\gamma^7\gamma_a\vartheta-\ups\gamma_a\gamma^7\ups)\,\Theta\Theta
+i(e^{a'}-\frac{1}{R}e^a\,\Theta\gamma^{a'}\gamma_a\Theta)\,\Theta\gamma_{a'}\gamma^7D\Theta
\nonumber\\
&&{}
+\frac{2i}{R^2}(e^{a'}-\frac{4}{3R}e^c\,\Theta\gamma^{a'}\gamma_c\Theta)\,e^a\,\ve_{abc}\,\Theta\gamma_{a'}\gamma^b\Theta\,\ups\gamma^c\ve\ups
+\frac{2}{R^2}(e^{a'}-\frac{2}{3R}e^b\,\Theta\gamma^{a'}\gamma_b\Theta)e^a\,\vartheta\gamma^7\gamma_a\vartheta\,\vartheta\gamma_{a'}\ups
\nonumber\\
&&{}
+\frac{1}{R^2}(e^{a'}-\frac{2}{3R}e^b\,\Theta\gamma^{a'}\gamma_b\Theta)e^a\,\Theta\gamma_{a'}\gamma^7\gamma_a\Theta\,(\vartheta\vartheta-\ups\ups)
-\frac{2}{3R^3}e^be^a\,\ve_{abc}\,\Theta\gamma^c\gamma^7\Theta\,((\vartheta\vartheta)^2-(\ups\ups)^2)
\nonumber\\
&&{}
+\frac{4i}{3R^3}e^be^d\,\vartheta\gamma_d\gamma^7\vartheta\,\vartheta\gamma^a\gamma^7\vartheta\,\ve_{abc}\,\ups\gamma^c\ve\ups
\Big)\,.
\eee
Note that the maximum order of the fermions in the above expressions
is ten.

Using the form of $F_4$ in (\ref{f4h3}) as well as the expressions
(\ref{A2-}) for ${\mathcal A}_1$ and (\ref{iH3-}) for $i_\Theta H_3$
the quantity relevant for computing the RR three-form potential
$\mathcal A_3$ becomes
\bee
\lefteqn{i_\Theta F_4+\mathcal A_1i_\Theta H_3}
\nonumber\\
&=&
-(e^b+i\Theta\gamma^bD\Theta+\frac{1}{R^2}e^b(\vartheta\vartheta-\ups\ups)^2)
(e^a+i\Theta\gamma^aD\Theta+\frac{1}{R^2}e^a(\vartheta\vartheta-\ups\ups)^2)\ve_{abc}
i\Theta\gamma^cD\Theta
\nonumber\\
&&{}
-2e^3(e^{a'}-\frac{2}{R}e^a\,\Theta\gamma^{a'}\gamma_a\Theta)D\Theta\gamma_{a'}\Theta
-\frac{4}{R}e^3e^a\Theta\gamma_aD\Theta\,\Theta\Theta(1+\frac{2}{R^2}(\ups\ups)^2)
\\
&&{}
\hspace{-10pt}-\frac{4}{R}(e^{a'}-\frac{2}{R}e^c\,\Theta\gamma^{a'}\gamma_c\Theta)e^a
(e^b+i\Theta\gamma^bD\Theta+\frac{1}{R^2}e^b(\vartheta\vartheta-\ups\ups)^2)
\Theta\gamma_{a'}\gamma_{ab}\Theta(1+\frac{2}{R^2}(\ups\ups)^2)
\nonumber\\
&&{}
\hspace{-10pt}-i(e^{b'}-\frac{2}{R}e^b\,\Theta\gamma^{b'}\gamma_b\Theta)(e^{a'}-\frac{2}{R}e^a\,\Theta\gamma^{a'}\gamma_a\Theta)D\Theta\gamma_{a'b'}\Theta
+kl_p(A(y)-\frac{2}{R^2}e^a\,\Theta\gamma^7\gamma_a\Theta)i_\Theta
H_3\,.\nonumber
\eee
One can now substitute this together with the expression for
$i_\Theta H_3$ (\ref{iH3-}) into eq. (\ref{A3}) and compute the
explicit form of the RR three--form potential ${\mathcal A}_3$ in
this gauge. Since we have not got a reasonably simple expression for
${\mathcal A}_3$ we shall not present it here.

\setcounter{equation}0
\section{Applications}
We can now use the kappa--gauge fixed form of the $AdS_4 \times
CP^3$ superbackground of Subsections \ref{theta-} and \ref{theta++}
to simplify the actions for the type IIA superstring and D--branes.
Let us note that the gauge fixing conditions (\ref{kappagauge1}) can
also be used to simplify the actions for the $D=11$ superparticle,
M2-- and M5--branes in the $AdS_4\times S^7/Z_k$ superbackground
(\ref{notaC}). We shall consider the example of the $D=11$ superparticle
below.

\subsection{$D=11$ superparticle}
\def\cm{\mathcal M}

Let us consider a massless superparticle in the $AdS_4\times
S^7/Z_k$ supergravity background. Recall that when $k=1,2$, the
supergravity background preserves the maximum number of 32
supersymmetries, while for $k>2$ it preserves only 24.
The superparticle action in the complete superspace
with 32 $\Theta$ is constructed using the supervielbeins of the
$OSp(8|4)/SO(7)\times SO(1,3)\times Z_k$ supercoset derived in
\cite{Gomis:2008jt}
\be\label{upsilonfunctions}
\begin{aligned}
\underline E^{\hat a}&=E^{\hat a}(x,y,\vartheta) + 4 i
\upsilon\gamma^{\hat a}\,{{\sinh^2{{\mathcal M}/ 2}}\over{\mathcal M}^2}\,D\upsilon
+\frac{R}{k}\,dz\,E_7{}^{\hat a}(\upsilon)\\
\underline E^{a'}&=E^{a'}(x,y,\vartheta)+2i\upsilon\,{{\sinh m}\over m}\gamma^{a'}\gamma^5\,E(x,y,\vartheta)\\
\underline E^7&=\frac{R}{k}\,dz\,\Phi(\upsilon)+
R\,\left(A(x,y,\vartheta)-\frac{4}{R}\,\upsilon\,\ve\gamma^5\,{{\sinh^2{{\mathcal
M}/2}}\over{\mathcal M}^2}\,D\upsilon\,\right) \\
\underline E^{\alpha i}&=\left({{\sinh{\mathcal M}}\over{\mathcal
M}}\,(D\upsilon-\frac{2}{k}\,dz\,\ve\upsilon)\right)^{\alpha i}
\\
\underline E^{\alpha a'}&=E^{\alpha a'}(x,y,\vartheta)-\frac{8}{R}E^{\beta
a'}\left(\gamma^5\,\upsilon\,{{\sinh^2{{m}/2}}\over{m}^2}\right)_{\beta
i}\upsilon^{\alpha i}\,,
\end{aligned}
\ee
where $z$ is the 7th, $U(1)$ fiber, coordinate of
$S^7$, $D\upsilon$ has been given in (\ref{D}) and the eight fermionic
coordinates $\upsilon^{\alpha i}$ correspond to the eight
supersymmetries broken by orbifolding with $k>2$.

The explicit form of the fermionic supervielbeins in
(\ref{upsilonfunctions}) and of the connections on\\
$OSp(8|4)/SO(7)\times SO(1,3)\times Z_k$ are not required for the
construction of the Brink--Schwarz superparticle action but one
needs them for the construction of the pure--spinor superparticle
action in curved superbackgrounds, so we present also the form of
the spin--connection below.

The $SO(1,3)$ connection is
\be\label{ads4connection}
\underline \Omega^{\hat a\hat b}= \Omega^{\hat a\hat b}(x,y,\vartheta)
+\frac{8}{R} \upsilon\gamma^{\hat a\hat
b}\gamma^5\,{{\sinh^2{{\mathcal M}/ 2}}\over{\mathcal
M}^2}\,\left(D\upsilon-\frac{2}{k}dz\,\ve\upsilon\right)
\ee
and the $SO(7)$ connection is
\bee
\underline \Omega^{a'b'}
&=&
\Omega^{a'b'}(x,y,\vartheta)-\frac{1}{R}\,\underline E^7\,J^{a'b'}
-\frac{2}{R}\,\upsilon\,{{\sinh m}\over m}\gamma^{a'b'}\gamma^5E\,,
\nonumber\\
\label{so7connection}
\\
{\underline\Omega}^{a'7}
&=&
\frac{1}{R}\left(\underline E^{b'}-4i\upsilon\,{{\sinh m}\over m}\gamma^{b'}\gamma^5E\right)\,J_{b'}{}^{a'}\,.\nonumber
\eee
The functions and forms appearing in
(\ref{upsilonfunctions})--(\ref{so7connection}) are defined in
Appendix B.

The first order form of the action for the massless superparticle in
the $OSp(8|4)/SO(7)\times SO(1,3)\times Z_k$ superbackground is
\bee\label{11dsuper}
S =  \int d\tau \left(  P_{\underline A} \,{\underline
E}^{\underline A}_\tau +
\frac{{e}}{2} \, P_{\underline A} P_{\underline B}\, \eta^{\underline{AB}} \right)\,,
\eee
where  $ P_{\underline A}$ ($\underline A=0,1,\dots,10$) is the
particle momentum, $e(\tau)$ is the Lagrange multiplier which
ensures the mass shell condition $P^2=0$ and
$$
{\underline E}^{\underline A}_\tau=\partial_\tau
Z^{\underline{\mathcal M}}\,{\underline E}_{\underline{\mathcal
M}}{}^{\underline A}\,, \qquad Z^{\underline{\mathcal
M}}=(x,y,z,\vartheta,\upsilon)
$$
is the pullback to the worldline of the supervielbeins
(\ref{upsilonfunctions}). The action is
invariant under local worldline diffeomorphisms and under the
fermionic kappa--symmetry transformations
\be\label{kappaA}
\delta Z^{\underline{\mathcal M}} \,{\underline E}_{\underline{\mathcal M}}{}^{\underline \alpha}
= P^{\underline A} \,(\Gamma_{\underline A}\,\kappa)^{\underline\alpha}\,,\qquad
\delta Z^{\underline{\mathcal M}} \,{\underline E}_{\underline
{\mathcal M}}{}^{\underline A} = 0\,, \qquad
\ee
\be\label{kappaA1}
\delta e=-4i\,{\underline E}^{\underline
\alpha}_\tau\,\kappa_{\underline\alpha}\,,
\qquad \delta P_{\underline A}=
\delta Z^{\underline{\mathcal M}}\,\underline\Omega_{{\underline{\mathcal M}}\underline A}{}^{\underline B}\,P_{\underline B}.
\ee
Inserting in the action the expressions for the vielbeins
(\ref{upsilonfunctions}), we get
\begin{eqnarray}\label{11dsuper2}
S =  \int d\tau  \!\!\!\!\!\!&&  \left[
P_{\hat a}\left(
E^{\hat a}_\tau + 4 i
\upsilon\gamma^{\hat a}\,{{\sinh^2{{\mathcal M}/ 2}}\over{\mathcal M}^2}\,D_\tau\upsilon
- 8 i \upsilon\gamma^{\hat a}\,{{\sinh^2{{\mathcal M}/2}}\over{\mathcal M}^2} \varepsilon {\upsilon} \,
\frac{\partial_\tau z}{k}
\right) \right. \nonumber \\
&&\left.+ P_{a'}
\left(E^{a'}_\tau+2i\upsilon\,{{\sinh m}\over m}\gamma^{a'}\gamma^5\,E_\tau \right)\right. \\
&&\left. + P_7 \left( R \left( \frac{\partial_\tau z}{k} + A\right)
-4
\upsilon \ve\gamma^5\,{{\sinh^2{{\mathcal M}/2}}\over{\mathcal
M}^2}\, (D_\tau \upsilon\, -2 \ve\upsilon \frac{\partial_\tau z}{k})
\right) +
\frac{{e}}{2} \, P_{\underline A} P_{\underline B}\, \eta^{\underline{AB}} \right]
\,.
\nonumber
\end{eqnarray}

The action (\ref{11dsuper2}) can be simplified by eliminating some
or all pure--gauge fermionic modes using the kappa--symmetry
transformations (\ref{kappaA}). For instance, when the momentum of
the particle is non--zero along a $CP^3$ direction inside $S^7$, the
projectors ${\mathcal P}_6$ and ${\mathcal P}_2$, defined in eqs.
(\ref{p6}) and (\ref{p2}), do not commute with the kappa--symmetry
projector (\ref{kappaA}) and one can use \emph{e.g.} the 16
kappa--symmetry transformations to eliminate 16 of the 24
$\vartheta$. After such a gauge fixing the action will contain 8
remaining $\vartheta$ and 8 $\upsilon$.

Alternatively, by partially gauge fixing the kappa--symmetry one can
eliminate all eight $\upsilon$ keeping 24 $\vartheta$. In the latter
case the action reduces to the form in which it describes the
dynamics of a superparticle in a superspace with 11 bosonic
coordinates and 24 fermionic ones. This superspace has been
introduced in
\cite{Gomis:2008jt} as a Hopf fibration of the supercoset $OSp(6|4)/U(3)\times SO(1,3)$.
It is the supercoset
\begin{equation}\label{24th}
\frac{OSp(6|4) \times U(1)}{U(3) \times SO(1,3)\times
Z_k}.
\end{equation}
The geometry of (\ref{24th}) is described by the supervielbeins
\begin{eqnarray}\label{24thA}
&&\hat E^{\hat a} = E^{\hat a}(x,y,\vartheta)\,, \nonumber \\
&&\hat E^{a'} = E^{a'}(x,y,\vartheta) \,,\\
&&\hat E^{7} = R\Big(\frac{dz}{k} + A(x,y,\vartheta) \Big)\,,
\nonumber \\
&&
\hat E^{\alpha a'} = E^{\alpha a'}(x,y,\vartheta)\,,\nonumber
\end{eqnarray}
where (as already mentioned) the explicit form of the
right--hand sides of (\ref{24thA}) are given in (\ref{cartan24}). Notice that now $z$  appears only in the vielbein $\hat
E^7$ along the $U(1)$-fiber direction of $S^7$.

The first order form of the superparticle action in the superspace
(\ref{24thA}) is
\bee\label{11dsuper24_A}
S& = & \int d\tau \left(  P_{\hat a} \,{\hat E}^{\hat a}_\tau +
 P_{a'} \,{\hat E}^{a'}_\tau +
  P_{7} \,{\hat E}^{7}_\tau +
\frac{{e}}{2} \, P_{\underline A} P_{\underline B}\, \eta^{\underline{AB}} \right)
\,\nonumber\\
& =&  \int d\tau \left(  P_{\hat a} \,{E}^{\hat a}_\tau +
 P_{a'} \,{E}^{a'}_\tau +
  P_{7} \, R \left( \frac{\partial_\tau z}{k} + A_\tau\right) +
\frac{{e}}{2} \, P_{\underline A} P_{\underline B}\, \eta^{\underline{AB}} \right)\,,
\eee
where now
$$
{\hat E}^{\underline A}_\tau=\partial_\tau Z^{{\mathcal M}}\,{\hat
E}_{{\mathcal M}}{}^{\underline A}+\partial_\tau\,z\,{\hat
E}_z{}^{\underline A}\,,
\qquad Z^{{\mathcal M}}=(x,y,\vartheta)
$$
is the pullback to the worldline of the supervielbeins
(\ref{24thA}).

It is easy to reduce the action (\ref{11dsuper24_A}) to $D=10$. Once it
is done, one obtains the action for a D0--brane moving in the
supercoset $OSp(6|4)/U(3)\times SO(1,3)$.

As we have mentioned above, the action (\ref{11dsuper24_A})
describes a superparticle which has a non--zero momentum along the
$CP^3$ base of the $S^7$ bundle. This is required by the consistency
of the kappa--symmetry gauge fixing condition $\upsilon=0$. To
describe other possible classical motions of the superparticle,
\emph{ e.g.} when $P^{a'}=0$, one should chose a different
kappa--symmetry gauge.

For instance, if the superparticle has a non--zero spacial momentum
along the 7th, fiber direction, of $S^7$, one can use the gauge
fixing condition corresponding to that of Subsection
\ref{theta-}. In this case, in virtue of the gauge-fixed expressions of Appendix C,
 the action (\ref{11dsuper2}) simplifies to
\begin{eqnarray}\label{11dsuper-gauged}
S =  \int d\tau \hspace{-.5cm}&&\left[
P_{a} \,\Big( e^a_\tau(x)+i\vartheta\gamma^aD_{\tau}\vartheta +
i \ups \gamma^aD_{\tau} \ups - 2 i
\ups \gamma^a \varepsilon \ups \frac{\partial_\tau z}{k} \Big) \right. \nonumber \\
&& \left. +P_{a'} \, {e}^{a'}_\tau(y) + P_{3} \,{e}^{3}_\tau(x) +
P_{7}
\, R
\left(\frac{\partial_\tau z}{k}  +  A_\tau(y)\right) +
\frac{{e}}{2} \, P_{\underline A} P_{\underline B}\, \eta^{\underline{AB}} \right]\,.
\end{eqnarray}
The dimensional reduction of the $D=11$ superparticle action
(\ref{11dsuper-gauged}) along $z$ results in the kappa--symmetry
gauge--fixed action which describes an arbitrary motion of the type
IIA D0--brane in $AdS_4\times CP^{3}$ superspace.

Before considering the D0--brane, let us note that the action
(\ref{11dsuper2}) is the most appropriate starting point for the
construction of the pure--spinor formulation of the $D=11$
superparticle in the $AdS_4\times CP^3$ supergravity background. The
pure--spinor condition $\lambda\Gamma^{\underline A}\lambda=0$ in
$D=11$ implies that the 32--component bosonic pure spinor
$\lambda^{\underline \alpha}$ has 23 independent components
\cite{Berkovits:2002uc,Fre':2006es}. This counting ensures the correct
number of bosonic and fermionic degrees of freedom.

In the cases of the actions (\ref{11dsuper24_A}) and
(\ref{11dsuper-gauged}) that describe a particle motion in the
reduced superspaces, one can also develop pure--spinor formulations
in which the pure spinor $\lambda$, in addition, is subject to the
same constraint as the one imposed on $\Theta$ by kappa--symmetry
gauge fixing, e.g. ${\mathcal P_2}\lambda=0$ in the case
$\upsilon={\mathcal P_2}\Theta=0$. This guarantees the correct
counting of the degrees of freedom in the pure--spinor formulation
(similar to the cases considered in
\cite{Fre:2008qc,Bonelli:2008us}). That is, the difference between
the bosonic and fermionic degrees of freedom remains the same.
Indeed, in the case of the pure--spinor formulation of the massless
$D=11$ superparticle
\cite{Berkovits:2002uc} there are 11 bosonic $X^{\underline A}$ plus 23 pure spinor
degrees of freedom and 32 fermionic $\Theta$, while  in the above
example of the reduced pure spinor formulation the pure spinor
effectively contains $23-8=15$ degrees of freedom against 24
fermionic ones, while the number of $X$ remains the same.

 When the pure spinor formulations of the superparticle
in reduced superspaces correspond to the kappa--gauge fixed versions
of the Brink--Schwarz superparticle whose consistency is limited to
particular subsectors of the classical configuration space of the
full theory, one may expect that the former will also describe only
subsectors of the pure--spinor superparticle model formulated in the
complete superspace with 32 fermionic coordinates. As in the case of
the pure--spinor type IIA superstring in $AdS_4\times CP^3$
\cite{Fre:2008qc,Bonelli:2008us}, these issues require additional
analysis.

\subsection{$D0$-brane}
To obtain the action for the $D0$--brane by dimensional
reduction of the $D=11$ superparticle action, one should first perform the
appropriate Lorentz transformation of the $D=11$
supervielbeins (as was explained in \cite{Gomis:2008jt}) and make a
corresponding redefinition of the particle momentum. We shall not
perform this dimensional reduction procedure since the result is
well known. The $D0$--brane action has the following first order
form in the type IIA superbackground in the string frame (see eqs.
(\ref{simplA}) and (\ref{simplB}))
\begin{equation}\label{d0-action}
S =  \int d\tau e^{- \phi}  \,\left(P_A {\cal E}^A_\tau +
\frac{e}{2} (P_A P_B \eta^{AB}+ m^2) \right) + m\,  \int {\cal
A}_1\,,
\end{equation}
where $m$ is the mass of the particle and the second term describes
its coupling to the RR one--form potential ${\cal A}_1$.

Integrating out the momenta $P_A$ and the auxiliary field $e(\tau)$
we arrive at the  action
\begin{equation}\label{d0-action2}
S = -m\,\int d\tau e^{- \phi}  \,\sqrt{-{\cal E}^A_\tau {\cal
E}^B_\tau\,\eta_{AB}} + m\int {\cal A}_1\,.
\end{equation}
The action (\ref{d0-action2}) is invariant under worldline
diffeomorphisms and the kappa-symmetry transformations (to
verify the kappa-symmetry one needs the superspace constraints
on the torsion $T^A$ and on $F_2$ given in Appendix A.4)
\begin{eqnarray}
&&\delta_\kappa Z^{\mathcal M} {\cal E}_{\cal
M}^{~~\underline\alpha} =
\frac{1}{2}(1 + \Gamma)^{\underline\alpha}{}_{\underline\beta} \,\kappa^{\underline\beta}(\tau)\,, ~~~
{\underline\alpha}  = 1,\dots,32\,,\qquad Z^{\mathcal M}=(x,y,\vartheta,\upsilon)\,\nonumber\\
\\
&&\delta_\kappa Z^{\mathcal M} {\cal E}_{\cal M}^{~~A} = 0\,, ~~~
A=0,1,\dots,9\,
\nonumber
\end{eqnarray}
where
\be\label{GammaD0}
\Gamma = \frac{1}{\sqrt{-\mathcal E^2_\tau}}\,{\mathcal E}_\tau^{~A} \Gamma_A \Gamma_{11}\,,
\,\qquad \Gamma^2=1\,.
\ee
Comparing the form of the kappa--symmetry projector matrix
(\ref{GammaD0}) with the kappa--symmetry gauge fixing condition of
Subsection \ref{theta-}, we see that
$\gamma=\Gamma^0\Gamma^1\Gamma^2$ (introduced in eq. (\ref{gamma}))
does not commute with $\Gamma$ in (\ref{GammaD0}) provided that the
energy $P^0\sim{{\mathcal E}^0_\tau}$ of the massive particle is
nonzero, which is always the case. Thus to simplify, \emph{e.g.} the first
order action (\ref{d0-action}) we can use the gauge fixed form of
the supervielbeins and the RR one--form of Subsection
\ref{theta-}. The action takes the following explicit form, with an
appropriately rescaled Lagrange multiplier $e(\tau)$,
\bee\label{simply}
S &=& \left(\frac{R}{kl_p}\right)^{-1} \,\int d\tau \,
\left[\Big( P_{a'}e_\tau^{a'}(y)
+ P_{3} \,e_\tau^3(x)\Big)\,(1+\frac{6}{R^2}(\ups\ups)^2)
\right. \nonumber \\
&+& \left. P_{a} \,
(e_\tau^a(x)+i\Theta\gamma^aD_{\tau}\Theta)\,(1+\frac{8}{R^2}(\ups\ups)^2)
+\frac{e(\tau)}{2} \,(P_A P_B \eta^{AB}+ m^2) \nonumber\right] \\
&+& m kl_p\,\int \, \Big(
A(y)-\frac{2i}{R^2}(e^a(x)+i\Theta\gamma^aD\Theta)\,
\ups\ve\gamma_a\ups\Big)\,.
\eee
This action contains fermionic terms up to the 6th order in
$\Theta=(\vartheta,\upsilon)$.

\subsection{The fundamental string}\label{FS}

In this section we use the geometry discussed above to construct
the Green-Schwarz model for the fundamental string. We will first
review the form of the superstring sigma model without gauge fixing
and then impose the gauge fixing of the kappa--symmetry. This will
provide a calculable sigma model.

The action for the Green--Schwarz superstring has the following form
\begin{equation}\label{cordaA}
S = -\frac{1}{4\pi\alpha'}\,\int d^2\xi\, \sqrt {-h}\, h^{IJ}\,
{\cal E}_{I}^{A} {\cal E}_{J}^{B} \eta_{AB}
-\frac{1}{2\pi\alpha'}\,\int  B_2\,,
\end{equation}
where $\xi^I$  $(I,J=0,1)$ are the worldsheet coordinates,
$h_{IJ}(\xi)$ is a worldsheet metric and $B_2$ is the pull--back to
the worldsheet of the NS--NS 2--form.

The kappa--symmetry transformations which leave the superstring
action (\ref{cordaA}) invariant are

\begin{equation}\label{kappastring}
\delta_\kappa Z^{\mathcal M}\,{\mathcal E}_{\mathcal M}{}^{\underline \alpha}=
{1\over 2}(1+\Gamma)^{\underline \alpha}_{~\underline\beta}\,
\kappa^{\underline\beta}(\xi),\qquad {\underline \alpha}=1,\cdots, 32
\ee
\be\label{kA}
\hskip-2.5cm\delta_\kappa Z^{\mathcal M}\,{\mathcal E}_{\mathcal M}{}^A=0,
\qquad   A=0,1,\cdots,9
\end{equation}
where $\kappa^{\underline\alpha}(\xi)$ is a 32--component spinor
parameter, ${1\over 2}(1+\Gamma)^{\underline
\alpha}_{~\underline\beta}$ is a spinor projection matrix with
\be\label{gbs}
\Gamma={1\over {2\,\sqrt{-\det{g_{IJ}}}}}\,\epsilon^{IJ}\,{\mathcal
E}_{I}{}^A\,{\mathcal E}_{J}{}^B\,\Gamma_{AB}\,\Gamma_{11}, \qquad
\Gamma^2=1\,,
\ee
and the auxiliary worldsheet metric $h^{IJ}$ transforms as follows
\bee\label{deltah}
\lefteqn{\delta_\kappa\,(\sqrt{-h}\,h^{IJ})}
\nonumber\\
&=&2i\,\sqrt{-h}\,(h^{IJ}\,g^{KL}-2h^{K(I}\,g^{J)L})
\left(\delta_\kappa
Z^{\mathcal M}\,{\mathcal E}_{\mathcal M}\,\Gamma_A\,{\mathcal
E}_K\,{\mathcal E}^A_L +\frac{1}{2}g_{KL}\delta_\kappa Z^{\mathcal
M}\,{\mathcal E}_{\mathcal M}{}^{\alpha i}\lambda_{\alpha i}
\right)
\\
&&\hspace{-20pt}
-2i\,\sqrt{-h}\,\,\frac{h^{IK'}g_{K'L'}h^{L'J}-\frac{1}{2}h^{IJ}\,h^{K'L'}g_{K'L'}}
{\frac{1}{2}\,h^{K'L'}\,g_{K'L'}+\sqrt{\frac{g}{h}}}\,g^{KL}\,\left(\delta_\kappa
Z^{\mathcal M}\,{\mathcal E}_{\mathcal M}\,\Gamma_A\,{\mathcal
E}_K\,{\mathcal E}^A_L +\frac{1}{2}g_{KL}\delta_\kappa Z^{\mathcal
M}\,{\mathcal E}_{\mathcal M}{}^{\alpha i}\lambda_{\alpha i}
\right)\nonumber
\eee
where
\be\label{im}
g_{IJ}(\xi)={\mathcal E}_{I}{}^{A}\, {\mathcal
E}_{J}{}^{B}\,\eta_{AB}\,,\qquad g^{IJ}\equiv (g_{IJ})^{-1}
\ee
is the induced metric on the worldsheet of the string that on the
mass shell coincides with the auxiliary metric $h_{IJ}(\xi)$ modulo
a conformal factor. Finally, $g=\det g_{IJ}$ and  $h=\det h_{IJ}$.
\\
Using the identity
\be
h^{IJ}\,g_{JK}\,h^{KL}\,g_{LI}-\frac{1}{2}\,(h^{IJ}\,g_{IJ})^{2}\equiv
\frac{1}{2}\,(h^{IJ}\,g_{IJ})^{2}-2\,\frac{g}{h}\,
\ee
one can check that eq. (\ref{deltah}) multiplied by $g_{IJ}$ results
in
\be\label{kappag}
\delta_\kappa\,(\sqrt{-h}\,h^{IJ})\,g_{IJ}
=4i(\sqrt{-g}\,g^{KL}-\sqrt{-h}\,h^{KL})\,
\delta_\kappa Z^{\mathcal M}\,{\mathcal E}_{\mathcal M}{}^{\underline\alpha}(\Gamma_A\,{\mathcal E}_K\,{\mathcal E}^A_L
+\frac{1}{2}g_{KL}\lambda )_{\underline\alpha}\,,
\ee
which together with the variation (\ref{kappastring}) and (\ref{kA})
of the superspace coordinates insures the invariance of the action
(\ref{cordaA}).

 Comparing the form of the kappa--symmetry projector
(\ref{kappastring}) with the kappa--symmetry gauge fixing condition
of Subsection
\ref{theta-} we see that this gauge choice is admissible when the
string moves in such a way that the projection of its worldsheet on
the $3d$ subspace along the directions $e^a$ $(a=0,1,2)$  of the
target space is a non--degenerate two--dimensional time--like
surface. Thus, it can be used to analyze the string dynamics in the
sector which is not reachable by the supercoset model of
\cite{Arutyunov:2008if,Stefanski:2008ik,Fre:2008qc}. The latter is
obtained from the action (\ref{cordaA}) by gauge fixing to zero the
eight fermions $\upsilon$, which is only possible when the string
worldsheet extends in the $CP^3$ directions.

In the gauge of Subsection \ref{theta-}  we insert into the action
(\ref{cordaA}) the  expressions (\ref{simple+v}) and (\ref{simple+B})
for the supervielbeins and $B_2$. This results in an action that
contains fermionic terms only up to the 8th order in
$\Theta=(\vartheta,\upsilon)$
\footnote{The factor in front of the
action is unconventional due to our normalization of the vielbeins,
which comes from the dimensional reduction of the eleven-dimensional
geometry. More conventional, unit radius string frame vielbeins
$(\hat e^{\hat a},\hat e^{a'})$ can be introduced by the following
rescaling
$$
(e^{\hat a},e^{a'})=\left(\frac{R}{k
l_p}\right)^{-1/2}\left(\frac{R}{2}\,\hat e^{\hat a},\,R\,\hat
e^{a'}\right)\,.
$$
Then the factor in front of the action becomes
$\frac{R^2}{4\pi\alpha'}=\frac{(R/l_p)^3}{4\pi k}$.}
\begin{eqnarray}\label{cordaB}
S &=&-\frac{1}{4\pi\alpha'}\,\frac{R}{k l_p}
\int\,d^2\xi\,\sqrt{-h}\,h^{IJ}\left[
\left(e^{a'}_I e^{b'}_{J} \delta_{a'b'} +
e^3_I e^3_{J}\right)\,(1-\frac{6}{R^2}(\ups\ups)^2)
\right.
\nonumber \\
&&{}
\hspace{4cm}
+\left.(e^a_I+i\Theta\gamma^aD_I\Theta)\,
(e^b_{J}+i\Theta\gamma^bD_{J}\Theta)\,\eta_{ab}\,
(1-\frac{2}{R^2}(\ups\ups)^2)\right]
\nonumber\\
\\
 &-&\frac{1}{2\pi\alpha'}\frac{R}{kl_p}\int
\Big[e^3\,\Theta\gamma^7D\Theta
+ie^{a'}\,\Theta\gamma_{a'}\gamma^7D\Theta
-\frac{2i}{R}e^3e^{a'}\,\vartheta\gamma_{a'}\gamma^7\ups
-\frac{1}{R}e^{b'}e^{a'}\,\Theta\gamma_{a'b'}\gamma^7\ups
\nonumber\\
&&{\hspace{60pt}}-\frac{i}{R}(e^b+i\Theta\gamma^bD\Theta)\,(e^a+i\Theta\gamma^aD\Theta)\,
\ups\gamma_{ab}\,\ve\ups\,
\Big]\,.\nonumber
\end{eqnarray}
To avoid possible confusion, let us remind the reader that in eqs.
(\ref{cordaB})--(\ref{cordaBT}) the covariant derivative
$D\Theta\equiv (D_8\upsilon,\,D_{24}\vartheta)$ is defined in eqs.
(\ref{D81}). Actually, in (\ref{cordaB}) the vielbein
$e^3$ does not contribute to the covariant derivative and the
connection $\omega^{ab}$ is zero along the $3d$ Minkowski boundary
of $AdS_4$, for the vielbeins chosen as in eqs. (\ref{ad4v}). It is
not hard to check that the action (\ref{cordaB}) is invariant
under twelve `linearly realized' supersymmetry transformations
$$
\delta\vartheta=\epsilon, \qquad \delta
e^a=-i\epsilon\,\gamma^a\,D_{24}\vartheta
$$
with parameters $\epsilon=\frac{1}{2}\,(1+\gamma)\,\epsilon$ being
$CP^3$ Killing spinors
$$
D_{24}\epsilon={\mathcal P}_6\,(d-\frac{1}{R}e^3
+\frac{i}{R}e^{a'}\gamma_{a'} -\frac{1}{4}\omega^{ab}\gamma_{ab}
-\frac{1}{4}\omega^{a'b'}\gamma_{a'b'})\epsilon=0\,.
$$
The other twelve supersymmetries of the $OSp(6|4)$ isometries of
(\ref{cordaB}) are non--linearly realized on the worldsheet fields
and include compensating kappa--symmetry transformations required to
maintain the gauge $\vartheta=\frac{1}{2}\,(1+\gamma)\,\vartheta$.

The action (\ref{cordaB}) is slightly more complicated than the
action for the $AdS_5\times S^5$ superstring in the analogous
kappa--symmetry gauge
\cite{Pesando:1998fv,Kallosh:1998nx,Kallosh:1998ji},
that contains fermions only up to the fourth order, since
$AdS_4\times CP^3$ is less supersymmetric than $AdS_5\times
S^5$. The action (\ref{cordaB}) takes a form similar to that of
\cite{Pesando:1998fv,Kallosh:1998nx,Kallosh:1998ji} when we formally
put the broken supersymmetry fermions $\upsilon^{\alpha i}$ to zero.

As in the case of the string in $AdS_5\times S^5$ it is possible to simplify the action further
by performing a T--duality transformation on the worldsheet \cite{Kallosh:1998ji}. Following
\cite{Kallosh:1998ji} we first rewrite the part of the action
(\ref{cordaB}) containing the vielbeins $e^a$ in the first order
form
\be\label{first}
S_1=\frac{1}{2\pi\alpha'}\,\frac{R}{k l_p}
\int\,d^2\xi\,\left[P_a^I\,(e^a_I+i\Theta\gamma^aD_I\Theta)
+\frac{1-\frac{6}{R^2}\,(\upsilon\upsilon)^2}{2\sqrt{-h}}\,P_a^I\,P_b^J\,h_{IJ}\,\eta^{ab}
-\frac{i}{R}\,\frac{\varepsilon_{IJ}}{-h}\,P^I_a\,P^J_b\,\ups\gamma^{ab}\ve\ups\right]\,.
\ee
The equations of motion for the momenta $P_a^I$ imply that
\be\label{P}
P^I_a=-\sqrt{-h}\,(1-\frac{2}{R^2}(\upsilon\upsilon)^2)\,
\Big(h^{IJ}\eta_{ab}
+\frac{2i}{R\sqrt{-h}}\,\varepsilon^{IJ}\,\ups\gamma_{ab}\ve\ups\Big)
\,(e^b_J+i\Theta\gamma^bD_J\Theta)\,.
\ee
Using the explicit form of the $AdS_4$ vielbeins given in eq.
(\ref{ad4v}) and varying the first order action (\ref{first}) with
respect to $x^a$ we find that $P^I_a$ is proportional to the
conserved current associated with translations along $x^a$
\be\label{trcurrent}
\partial_I\,\Big({r^2\over R^2}\,P^I_a\Big)=0\, \qquad\Rightarrow
\qquad P^I_a=\frac{R^2}{r^2}\,\varepsilon^{IJ}\partial_J\,{\tilde
x}_a\equiv \varepsilon^{IJ}\,{\tilde e}_{Ja}\,.
\ee
If we now substitute eq. (\ref{trcurrent}) into (\ref{first}) the
T--dualized version of the action (\ref{cordaB}) for the string in
$AdS_4\times CP^3$ takes the form
\begin{eqnarray}\label{cordaBT}
S &=&-\frac{1}{4\pi\alpha'}\,\frac{R}{k l_p}
\int\,d^2\xi\,\sqrt{-h}\,h^{IJ}
\left(
{\tilde e}_I^a\,{\tilde e}_J^b\,\,\eta_{ab}+e^3_I e^3_{J}
+e^{a'}_I e^{b'}_{J} \delta_{a'b'}
\right)\,(1-\frac{6}{R^2}(\ups\ups)^2)
\nonumber\\
\\
&-&\frac{1}{2\pi\alpha'}\frac{R}{kl_p}\int
\Big[e^3\,\Theta\gamma^7D\Theta
+ie^{a'}\,\Theta\gamma_{a'}\gamma^7D\Theta
-\frac{2i}{R}e^3e^{a'}\,\vartheta\gamma_{a'}\gamma^7\ups
-\frac{1}{R}e^{b'}e^{a'}\,\Theta\gamma_{a'b'}\gamma^7\ups
\nonumber\\
&&{\hspace{60pt}} +i{\tilde e}^a\,\Theta\gamma_aD\Theta
-\frac{i}{R}\,\,{\tilde e}^a\,{\tilde e}^b\,\ups\gamma_{ab}\ve\ups
\Big]\,.\nonumber
\end{eqnarray}

Note that in the T--dualized action the fermionic kinetic terms
appear only in the Wess--Zumino term and that there are now terms of
at most fourth order in fermions. Note also that the first (induced
metric) term of (\ref{cordaBT}) acquires a common factor
$(1-\frac{6}{R^2}(\ups\ups)^2)$ in contrast to the corresponding
terms in the original action (\ref{cordaB}).

To preserve the conformal invariance of the dual action at the
quantum level one should add to it a dilaton term
$\int\,R^{(2)}\,\tilde
\phi$ (where $R^{(2)}$ is the worldsheet curvature), which is
induced by the functional integration of $P_a^I$ when passing to the
dual action (see \cite{Buscher:1987qj,Schwarz:1992te,Kallosh:1998ji}
for details). Here we should point out that in our case the original
$AdS_4\times CP^3$ superbackground already has a non--trivial
dilaton which depends on $\upsilon$ (see eqs. (\ref{dilaton1}) and
(\ref{gfdialton})).

The following comment is now in order. As in the $AdS_5\times S^5$
case \cite{Kallosh:1998ji,Ricci:2007eq,Beisert:2008iq}, upon the
T--duality along the three translational directions $x^a$ of $AdS_4$
the purely bosonic (classically integrable) $AdS_4\times CP^3$
sector of the type IIA superstring sigma model maps into an
equivalent sigma model on a dual $AdS_4$ space, both models sharing
the same integrable structure \cite{Ricci:2007eq}. The situation
with the fermionic sector of the $AdS_4\times CP^3$ superstring is,
however, different due to the fact that there is less supersymmetry
than in the $AdS_5\times S^5$ case.

 In the case of the
$AdS_5\times S^5$ superstring sigma model, one can accompany the
above bosonic T--duality transformation by a fermionic one along
fermionic directions in (complexified) superspace which have
translational isometries \cite{Berkovits:2008ic,Beisert:2008iq}.
This compensates the dilaton term generated by the bosonic
T--duality and maps the $AdS_5\times S^5$ superstring action to an
equivalent (dual) one, which is also integrable\footnote{In
\cite{Ricci:2007eq,Berkovits:2008ic,Beisert:2008iq} it has been shown that this
duality property of the $AdS_5\times S^5$ superstring is related to
earlier observed dual conformal symmetry of maximally helicity
violating amplitudes of the ${\mathcal N}=4$ super--Yang--Mills
theory and to the relation between gluon scattering amplitudes and
Wilson loops at strong and weak coupling.}. However, in the
$AdS_4\times CP^3$ case under consideration the fermionic directions
in superspace parametrized by $\upsilon$ do not have translational
isometries, since the action (\ref{cordaB}) or (\ref{cordaBT}) has
$\upsilon$--dependent fermionic terms which do not contain
worldsheet derivatives. This just reflects the fact that the
fermionic modes $\upsilon$ correspond to the broken supersymmetries
of the superbackground.

As far as the T--dualization of the supersymmetric fermionic modes
$\vartheta$ is concerned, it might be, in principle, possible (at
least in the absence of $\upsilon$) if, as in the case of the
$PSU(2,2|4)$ superstring sigma model
\cite{Berkovits:2008ic,Beisert:2008iq}, there existed a
realization of the $OSp(6|4)$ superalgebra in which 12 of the 24
(complex conjugate) supersymmetry generators squared to zero and
formed a representation of the bosonic subalgebra of $OSp(6|4)$. In
other words the possibility of T--dualizing part of fermionic modes
$\vartheta$ (in the absence of $\upsilon$) is related to the
question of the existence of a chiral superspace representation of
the superalgebra $OSp(6|4)$. Such a realization of $OSp(6|4)$ seems
not to exist. In fact, it has been argued in \cite{Adam:2009kt} that
the $OSp(6|4)$ supercoset subsector of the Green--Schwarz
superstring in $AdS_4\times CP^3$  does not have any fermionic
T--duality symmetry since in $OSp(6|4)$ the dimension of the
representation of the supercharges under the R--symmetry is odd. The
absence of the fermionic T--duality of the superstring in
$AdS_4\times CP^3$ may have interesting manifestations in particular
features of the $AdS_4/CFT_3$ holography.

The gauge--fixed actions (\ref{cordaB}) or (\ref{cordaBT}) can be
used for studying different aspects of the $AdS_4/CFT_3$
correspondence and integrability on both of its sides, in
particular, for making two-- and higher--loop string computations
for testing the Bethe ansatz and the S--matrix
\cite{Minahan:2008hf}--\cite{Minahan:2009te} in the dual planar ${\mathcal N}=6$
superconformal Chern--Simons--matter theory, which would extend the
analysis of
\cite{Nishioka:2008gz}--\cite{Suzuki:2009sc},\cite{Zarembo:2009au} and
others.

\subsection{D2--branes}
Let us now consider the effective worldvolume theory of probe
D2--branes moving in the $AdS_4 \times CP^3$ superbackground. This
can be derived from the action for $D$--branes in a generic type IIA
superbackground
\cite{Cederwall:1996ri,Aganagic:1996pe,Bergshoeff:1996tu} by
substituting the explicit form of the $AdS_4 \times CP^3$
supergeometry (\ref{simplA})--(\ref{A3}).

The action for a D2--brane in a generic type IIA supergravity
background in the string frame has the following form
\begin{equation}\label{DBIstring}
S= -T\int\,d^{3}\xi\,e^{{ - } {\phi}}\sqrt{-\det(g_{IJ}+{\cal
F}_{IJ})}+T \int\, ({\mathcal A}_3 + {\mathcal A}_1\,{\cal F}_2)\,,
\end{equation}
where $T$ is the tension of the D2--brane, $\phi(Z)$ is the dilaton
superfield,
\be\label{imb}
g_{IJ}(\xi)={\mathcal E}_{I}{}^{A}\, {\mathcal
E}_{J}{}^{B}\,\eta_{AB}\qquad I,J=0,1,2;\qquad A,B=0,1,\cdots,9
\ee
is the induced metric on the D2--brane worldvolume with ${\mathcal
E}_{I}{}^{A}=\partial_I\,Z^{\mathcal M}\,{\mathcal E}_{\mathcal
M}{}^{A}$ being the pullbacks of the vector supervielbeins of the
type IIA $D=10$ superspace and
\begin{equation}\label{deltaQstring}
{\cal F}_2 = d{\mathcal V} - {B}_2
\end{equation}
is the field strength of the worldvolume Born--Infeld gauge field
${\mathcal V}_I(\xi)$ extended by the pullback of the NS--NS
two--form. ${\mathcal A}_1$ and  ${\mathcal A}_3$ are the pullbacks
of the type IIA supergravity RR superforms (\ref{simplB}) and
(\ref{A3}).

Provided that the superbackground satisfies the IIA supergravity
constraints, the action (\ref{DBIstring}) is invariant under
kappa--symmetry transformations of the superstring coordinates
$Z^{\mathcal M}(\xi)$ of the form
(\ref{kappastring}), (\ref{kA}), together with
\be
\delta_\kappa\mathcal F_{IJ}=-\mathcal E_J^{\mathcal B}\,\mathcal E_I^{\mathcal A}\,\delta_\kappa Z^{\mathcal M}\mathcal E_{\mathcal M}{}^{\underline\alpha}\,H_{\underline\alpha\mathcal{AB}}
=-4i\mathcal E_{[I}^A\,\mathcal E_{J]}\Gamma_A\Gamma_{11}\delta_\kappa\mathcal E
-2i\mathcal E_J^B\,\mathcal E_I^A\,\delta_\kappa\mathcal E\Gamma_{AB}\Gamma_{11}\lambda\,,
\ee
where $\delta_\kappa\mathcal E=\delta_\kappa Z^{\mathcal M}\mathcal E_{\mathcal M}{}^{\underline\alpha}$.

In the case of the $D2$--brane the matrix $\Gamma$ has the form
\bee\label{bargamma}
\Gamma&=-&{{1}\over {\sqrt{-\det(g+{\cal
F})}}}\,\,\varepsilon^{IJK}\,({1\over 3!}\,{\mathcal
E}_I^{A}\,{\mathcal E}_J^{B}\,{\mathcal E}_K^{C}\Gamma_{ABC}+{1\over
2}\,{\mathcal F_{IJ}}\,{\mathcal
E}_K^{A}\,\Gamma_A\,\Gamma_{11})\nonumber\\
&&\\
&=&-{{1}\over {3!\sqrt{-\det(g+{\cal
F})}}}\,\,\varepsilon^{IJK}\,{\mathcal E}_I^{A}\,{\mathcal
E}_J^{B}\,{\mathcal E}_K^{C}\,\Gamma_{ABC}\,\,(1+{1\over
2}\,{\mathcal F^{IJ}}\,{\mathcal E}_I^{A}\,{\mathcal
E}_J^{B}\,\Gamma_{AB}\,\Gamma_{11})\,.\nonumber
\eee

\subsubsection{D2 filling $AdS_2\times S^1$ inside of
$AdS_4$}\label{d2ads2s1}

Let us consider the D2--brane configuration which corresponds to a
disorder loop operator in the ABJM theory
\cite{Drukker:2008jm}. The 1/2 BPS static solution of the equations of motion of the D2--brane on
$AdS_2\times S^1$ in the metric
\be\label{ads4}
ds^2=\frac{R_{CP^3}^2}{4u^2}(-dx^0\,dx^0+dr^2+r^2d\varphi^2+du^2)+R_{CP^3}^2ds^2_{CP^3}\,,
\ee
where
\bee
ds^2_{CP^3}={1\over 4}\Big[d\alpha^2+\cos^2{\alpha\over
2}(d\vartheta_1^2+\sin^2\vartheta_1d\varphi_1^2)+\sin^2{\alpha\over
2}(d\vartheta_2^2+\sin^2\vartheta_2d\varphi_2^2)\cr
+\sin^2{\alpha\over 2}\cos^2{\alpha\over 2}(d\chi+
\cos\vartheta_1d\varphi_1-\cos\vartheta_2d\varphi_2)^2\Big]\,,\nonumber
\eee
is characterized by the following embedding of the brane worldvolume
\be\label{ads2s11}
\xi^0=x^0,\qquad \xi_1=u,\qquad \xi_2=\varphi\,,
\qquad r=a\,\xi_1
\ee
which is supported by the non--zero (electric) Born--Infeld field
strength
\be
F=E {dx^0\wedge du\over u^2}\,,  \qquad
E=\frac{R_{CP^3}^2}{4}\sqrt{1+a^2}\,,
\ee
where $a$ is an arbitrary constant. Note that the presence of the
non--zero DBI flux on the $AdS_2$ subspace of the D2--brane
worldvolume is required to ensure the no--force condition, i.e.
vanishing of the classical action (\ref{DBIstring}) of this static
D2--brane configuration, provided that also an additional BI flux
boundary counterterm is added to the action (see
\cite{Drukker:2008jm} for more details). A natural explanation of
this boundary term is that it appears in the process of the
dualization of the compactified 11th coordinate scalar field of the
M2--brane into the BI vector field of the D2--brane.

Note that in \cite{Drukker:2008jm} this brane configuration was
considered in a different coordinate system, in which $AdS_4$ is
foliated with $AdS_2\times S^1$ slices instead of the flat $R^{1,2}$
slices. This makes manifest the symmetries of the D2--brane
configuration. An explicit form of the $AdS_4$ metric in this
slicing is
\be\label{ads2s1}
ds^2_{_{AdS_4}}= {R_{CP^3}^2\over 4}\,(\cosh^2 \psi
\,ds^2_{_{AdS_2}}+d\psi^2+
\sinh^2 \psi
\, d\varphi^2)
\ee
which is essentially a double analytic continuation of the usual
global $AdS_4$ metric. The static D2--brane configuration is then
characterized by the identification of the worldvolume coordinates
$\xi^a$ with those of $AdS_2$ and the $S^1$ angle $\varphi$.
However, for our choice of the kappa--symmetry gauge fixing
condition the use of the metric in the form (\ref{ads4}) is more
convenient, since the associated $AdS_4$ vielbeins
\be\label{0123}
e^0=\frac{R_{CP^3}}{2u}\,dx^0\,,\qquad
e^1=\frac{R_{CP^3}}{2u}\,dr\,,\qquad
e^2=\frac{R_{CP^3}\,r}{2u}\,d\varphi\,,\qquad
e^3=-\frac{R_{CP^3}}{2u}\,du\,
\ee
and the spin connection directly satisfy the relations
(\ref{eaoa31}) and (\ref{oab}).

One can be interested in D2--brane bosonic and fermionic
fluctuations around this 1/2 BPS static D2--brane solution described
by the action (\ref{DBIstring}). To simplify the form of the
fermionic terms, the kappa--symmetry gauge fixing for the D2--brane
wrapping $AdS_2
\times S^1$ can be made in the simplest possible way considered in
Subsection \ref{theta-}. To get the gauge fixed D2--brane action in
this case one should substitute into (\ref{DBIstring}) the
expressions for the vector supervielbeins (\ref{simple+v}), the RR
one--form (\ref{simple+A}) and the three--form (\ref{A3-}), and the
NS--NS two--form (\ref{simple+B}).

\subsubsection{D2 at the Minkowski boundary of $AdS_4$}
Let us now consider the supersymmetric effective worldvolume action
describing a D2--brane placed at the Minkowski boundary of the
$AdS_4$ space. In this case it is convenient to choose the
$AdS_4\times CP^3$ metric in the form (\ref{ads4metric11}) or
(\ref{ads4metric21}).

When the D2--brane is at the Minkowski boundary, we take the static
gauge $\xi^m=x^m$. The 1/2 BPS ground state of the D2--brane is when
its transverse scalar modes are constant and the Born--Infeld field
and the fermionic modes are zero. As a consistency check, let us
note that with the choice of the background value of the RR 3--form
(\ref{A3}) and (\ref{A31}) and of the corresponding (positive) $D2$--brane charge
(characterized by the plus sign in front of the Wess--Zumino term
(\ref{DBIstring})), the action of the ground state of the D2--brane
at the Minkowski boundary vanishes. This means that such a brane
configuration is stable and does not experience any external force,
\emph{i.e.} it is a BPS state.

If, on the other hand, with the same choice of ${\mathcal A}_3$
(\ref{A3}) and (\ref{A31}), we considered an anti--$D2$--brane carrying a negative
${\mathcal A}_3$ charge (which would be characterized by a minus
sign in front of the Wess--Zumino term in (\ref{DBIstring})), the
ground state of this anti--$D2$--brane at the Minkowski boundary
would have a non--zero action
$$
S_{\overline{D2}}=-2Te^{-\phi_0}\,\int\,d^{\,3}x\,\left(r\over {R_{CP^3}}\right)^6\,
$$
implying that such a solution is unstable (as is well known to be
the case for a probe anti--D--brane in a background of D--branes).
It is, therefore, important for the consistency of the solution to
take care that the relative signs of the RR potential ${\mathcal
A}_3$ and the $D2$--brane charge (and, as a consequence, the sign of
the kappa--symmetry projector) ensure the no--force condition, i.e.
vanishing of the static $D2$--brane action. In the case of M2, M5
and D3--branes at the Minkowski boundary of $AdS$ this issue was
discussed in detail in
\cite{Pasti:1998tc}.

For the static D2--brane configuration the kappa--symmetry projector
(\ref{bargamma}) reduces to
\be\label{kappa}
P=\frac{1}{2}(1+\gamma), \qquad
\gamma=\gamma^0\gamma^1\gamma^2=-\gamma_0\gamma_1\gamma_2
\ee
So the natural choice of the kappa--symmetry gauge fixing condition
is
\be\label{kappagauge}
\Theta=\frac{1}{2}(1-\gamma)\Theta\,,
\ee
\emph{i.e.} the gauge choice considered in detail in Subsection
\ref{theta++}.
Note that in the case of the D2--brane at the Minkowski boundary we
cannot use the simpler condition
$\Theta=\frac{1}{2}(1+\gamma)\Theta$ of Subsection \ref{theta-},
because the kappa--symmetry projector (\ref{kappa}) has the same
sign.

Plugging the kappa--symmetry gauge--fixed quantities of Subsection
3.2 into the action (\ref{DBIstring}), one can study the properties
of the $OSp(6|4)$ invariant effective $3d$ gauge--matter field
theory on the worldvolume of the $D2$--brane placed at the Minkowski
boundary of $AdS_4$, which from the point of view of M--theory
corresponds to an $M2$--brane pulled out to a finite distance from a
stack of $M2$--branes probing $R^8/Z_k$.

The effective theory on the worldvolume of this D2--brane, which
describes its fluctuations in $AdS_4
\times CP^3$, is an interacting $d=3$ gauge Born--Infeld--matter theory
possessing the (spontaneously broken) superconformal symmetry
$OSp(6|4)$. The model is superconformally invariant in spite of the
presence on the $d=3$ worldvolume of the dynamical Abelian vector
field, since the latter is coupled to the $3d$ dilaton field
associated with the radial direction of $AdS_4$. The superconformal
invariance is spontaneously broken by a non--zero expectation value
of the dilaton. An ${\mathcal N=3}$ superfield model with similar
symmetry properties was considered in the Appendix of
\cite{Buchbinder:2008vi}. To establish the explicit relation between
 the two models one should extract from the
superfield action of \cite{Buchbinder:2008vi} the component terms
describing its physical sector and compare the result with
corresponding terms in the D2--brane action.

\section{Conclusion}
In this paper we have considered the gauge--fixing of
kappa--symmetry of the superparticle, superstring and D2--brane
actions in the complete $AdS_4\times CP^3$ superspace which is
suitable, in particular, for studying regions of these theories that
are not reachable by partially kappa--symmetry gauge fixed models
based on the supercoset $OSp(6|4)/U(3)\times SO(1,3)$. The
simplified form of these actions can be used to approach various
problems of the $AdS_4/CFT_3$ correspondence. The gauge fixed form
of the $AdS_4\times CP^3$ supergeometry can also be used to consider
the actions for higher dimensional D4--, D6-- and D8--branes.

\section*{Acknowledgments}
The authors would like to thank Pietro Fr\'e and Jaume Gomis for
collaboration at early stages of this project and for many fruitful
discussions and comments. D.S. is also thankful to Soo--Jong Rey for
useful discussions. P.A.G. and D.S. are grateful to the Organizers
of the Workshop Program ``Fundamental Aspects of Superstring Theory"
for their hospitality at KITP, Santa Barbara, where their research
was supported in part by the National Science Foundation under Grant
No. PHY05-51164. Work of P.A.G., D.S. and L.W. was partially
supported by the INFN Special Initiative TV12. D.S. was also
partially supported by the INTAS Project Grant 05-1000008-7928, an
Excellence Grant of Fondazione Cariparo and the grant FIS2008-1980
of the Spanish Ministry of Science and Innovation.

\def\thesection{}
\def\theequation{A.\arabic{equation}}\label{A}
\section{Appendix A. Main notation and conventions}
\setcounter{equation}0

The convention for the ten and eleven dimensional metrics is the
`almost plus' signature $(-,+,\cdots,+)$. Generically, the tangent
space vector indices are labeled by letters from the beginning of
the Latin alphabet,  while  letters from the middle of the Latin
alphabet stand for curved (world) indices. The spinor indices are
labeled by Greek letters.

\def\thesubsection{A.1}
\subsection{$AdS_4$ space}

$AdS_4$ is parametrized by the coordinates $x^{\hat m}$ and its
vielbeins are $e^{\hat a}=dx^{\hat m}\,e_{\hat m}{}^{\hat a}(x)$,
${\hat m}=0,1,2,3;$ ${\hat a}=0,1,2,3$. The $D=4$ gamma--matrices
satisfy:
\be\label{gammaa}
\{\gamma^{\hat a},\gamma^{\hat b}\}=2\,\eta^{\hat a\hat b}\,,
\qquad \eta^{\hat a\hat b}={\rm diag}\,(-,+,+,+)\,,
\ee
\be\label{gamma5}
\gamma^5=i\gamma^0\,\gamma^1\,\gamma^2\,\gamma^3, \qquad
\gamma^5\,\gamma^5=1\,.
\ee
The charge conjugation matrix $C$ is antisymmetric, the matrices
$(\gamma^{\hat a})_{\alpha\beta}\equiv (C\,\gamma^{\hat
a})_{\alpha\beta}$ and $(\gamma^{\hat a\hat
b})_{\alpha\beta}\equiv(C\,\gamma^{\hat a\hat b})_{\alpha\beta}$ are
symmetric and $\gamma^5_{\alpha\beta}\equiv
(C\gamma^5)_{\alpha\beta}$ is antisymmetric, with
$\alpha,\beta=1,2,3,4$ being the indices of a 4--dimensional spinor
representation of $SO(1,3)$ or $SO(2,3)$.

\def\thesubsection{A.2}
\subsection{$CP^3$ space}

$CP^3$ is parametrized by the coordinates $y^{m'}$ and its vielbeins
are $e^{a'}=dy^{m'}e_{m'}{}^{a'}(y)$, ${m'}=1,\cdots,6;$
${a'}=1,\cdots,6$. The $D=6$ gamma--matrices satisfy:
\be\label{gammaa'}
\{\gamma^{a'},\gamma^{b'}\}=2\,\delta^{{a'}{b'}}\,,\qquad \delta^{a'b'}={\rm
diag}\,(+,+,+,+,+,+)\,,
\ee
\be\label{gamma7}
\gamma^7={i\over{6!}}\,\epsilon_{\,a_1'a_2'a_3'a_4'a_5'a_6'}\,\gamma^{a_1'}\cdots \gamma^{a_6'} \qquad
\gamma^7\,\gamma^7=1\,.
\ee
The charge conjugation matrix $C'$ is symmetric and the matrices
$(\gamma^{a'})_{\alpha'\beta'}\equiv
(C\,\gamma^{a'})_{\alpha'\beta'}$ and
$(\gamma^{a'b'})_{\alpha'\beta'}\equiv(C'\,\gamma^{a'b'})_{\alpha'\beta'}$
are antisymmetric, with $\alpha',\beta'=1,\cdots,8$ being the
indices of an 8--dimensional spinor representation of $SO(6)$.

\def\thesubsection{A.3}
\subsection{ Type IIA  $AdS_4\times CP^3$ superspace}

The type IIA superspace whose bosonic body is $AdS_4\times CP^3$ is
parametrized by 10 bosonic coordinates $X^M=(x^{\hat m},\,y^{m'})$
and 32-fermionic coordinates
$\Theta^{\underline\mu}=(\Theta^{\mu\mu'})$
($\mu=1,2,3,4;\,\mu'=1,\cdots,8$). These  combine into the
superspace supercoordinates $Z^{\cal M}=(x^{\hat
m},\,y^{m'},\,\Theta^{\mu\mu'})$. The type IIA supervielbeins are
\begin{equation}\label{IIAsv}
{\mathcal E}^{\mathcal A}=dZ^{\mathcal M}\,{\mathcal E}_{\mathcal
M}{}^{\mathcal A}(Z)=({\mathcal E}^{A},\,{\mathcal
E}^{\underline\alpha})\,,\qquad {\mathcal E}^{A}(Z)=({\mathcal
E}^{\hat a},\,{\mathcal E}^{a'})\,,\qquad {\mathcal
E}^{\underline\alpha}(Z)={\mathcal E}^{\alpha\alpha'}\,.
\ee
\def\thesubsection{A.4}
\subsection{Superspace constraints}
In our conventions the superspace constraint on the bosonic part of the torsion is
\be
T^A=-i\mathcal E\Gamma^A\mathcal E+i\mathcal E^A\,\mathcal
E\lambda+\frac{1}{3}{\mathcal E}^A\,\mathcal E^B\nabla_B\phi\,,
\ee
while the constraints on the RR and NS--NS field strengths are
\begin{eqnarray}
F_2&=&-i\,e^{-\phi}\,\mathcal E\Gamma_{11}\mathcal E
+2i\,e^{-\phi}\,\mathcal E^A\,\mathcal E\Gamma_A\Gamma_{11}\lambda+\frac{1}{2}\mathcal E^B\mathcal E^A\,F_{AB}\,,\\
F_4&=&-\frac{i}{2}\,e^{-\phi}\,{\mathcal E}^B{\mathcal E}^A\,\mathcal E\Gamma_{AB}\mathcal E
+\frac{1}{4!}{\mathcal E}^D{\mathcal E}^C{\mathcal E}^B{\mathcal E}^A\,F_{ABCD}
\,,\\
H_3&=&
-i{\mathcal E}^A\,\mathcal E\Gamma_A\Gamma_{11}\mathcal E
+i{\mathcal E}^B{\mathcal E}^A\,\mathcal E\Gamma_{AB}\Gamma^{11}\lambda
+\frac{1}{3!}{\mathcal E}^C{\mathcal E}^B{\mathcal E}^A\,H_{ABC}\,.
\end{eqnarray}
These differ from the conventional string frame constraints by the $\lambda$--term in $T^A$ and related terms in $F_2$, $F_4$ and $H_3$. This is
a consequence of the dimensional reduction from eleven dimensions. They can be brought to a more conventional form by shifting the fermionic supervielbein $\mathcal E^{\underline\alpha}$ by $-\frac{1}{2}\mathcal E^A(\Gamma_A\lambda)^{\underline\alpha}$ accompanied by a related shift in the connection.
\\
\\
{\bf The $D=10$ gamma--matrices $\Gamma^A$} are given by
\bee\label{Gamma10}
&\{\Gamma^A,\,\Gamma^B\}=2\eta^{AB},\qquad
\Gamma^{A}=(\Gamma^{\hat a},\,\Gamma^{a'})\,,\nonumber\\
&\\
&\Gamma^{\hat a}=\gamma^{\hat a}\,\otimes\,{\bf 1},\qquad
\Gamma^{a'}=\gamma^5\,\otimes\,\gamma^{a'},\qquad
\Gamma^{11}=\gamma^5\,\otimes\,\gamma^7,\qquad a=0,1,2,3;\quad
a'=1,\cdots,6\,. \nonumber
\eee
The charge conjugation matrix is ${\mathcal C}=C\otimes C'$.

The fermionic variables $\Theta^{\underline\alpha}$ of IIA
supergravity carrying 32--component spinor indices of $Spin(1,9)$,
in the $AdS_4\times CP^3$ background and for the above choice of the
$D=10$ gamma--matrices, naturally split into 4--dimensional
$Spin(1,3)$ indices and 8--dimensional spinor indices of $Spin(6)$,
i.e. $\Theta^{\underline\alpha}=\Theta^{\alpha\alpha'}$
($\alpha=1,2,3,4$; $\alpha'=1,\cdots,8$).

\def\thesubsection{A.5}
\subsection{$24+8$ splitting of $32$ $\Theta$}

24 of  $\Theta^{\underline\alpha}=\Theta^{\alpha\alpha'}$ correspond
to the unbroken supersymmetries of the $AdS_4\times CP^3$
background. They are singled out by a projector introduced in
\cite{Nilsson:1984bj} which is constructed using the $CP^3$ K\"ahler
form $J_{a'b'}$ and seven $8\times 8$ antisymmetric gamma--matrices
(\ref{gammaa'}). The $8\times 8$ projector matrix has the following
form
\be\label{p6}
{\mathcal P}_{6}={1\over 8}(6-J)\,,
\ee
where the $8\times 8$  matrix
\be\label{J}
J=-iJ_{a'b'}\,\gamma^{a'b'}\,\gamma^7 \qquad {\rm such~ that} \qquad
J^2= 4J+12
\ee
has six eigenvalues $-2$ and two eigenvalues $6$, \emph{i.e.} its
diagonalization results in
\be\label{Jdia}
J=\hbox{diag}(-2,-2,-2,-2,-2,-2,6,6)\,.
\ee
Therefore, the projector (\ref{p6}) when acting on an 8--dimensional
spinor annihilates 2 and leaves 6 of its components, while the
complementary projector
\be\label{p2}
{\mathcal P}_{2}={1\over 8}(2+J)\,,\qquad
\mathcal{P}_2+\mathcal{P}_6=\mathbf 1
\ee
annihilates 6 and leaves 2 spinor components.

Thus the spinor
\be\label{24}
\vartheta^{\alpha\alpha'}=({\mathcal P}_6\,\Theta)^{\alpha\alpha'} \qquad \Longleftrightarrow \qquad
\vartheta^{\alpha a'}\, \qquad a'=1,\cdots, 6
\ee
has 24 non--zero components and the spinor
\be\label{8}
\upsilon^{\alpha\alpha'}=({\mathcal P}_2\,\Theta)^{\alpha\alpha'}\qquad \Longleftrightarrow \qquad
\upsilon^{\alpha i}\, \qquad i=1,2
\ee
has 8 non--zero components. The latter corresponds to the eight
supersymmetries broken by the $AdS_4\times CP^3$ background.

To avoid confusion, let us note that the index $a'$ on spinors is
different from the same index on bosonic quantities. They are
related by the usual relation between vector and spinor
representations, \emph{i.e.} given two $Spin(6)$ spinors
$\psi_1^{\alpha'}$ and $\psi_2^{\alpha'}$, projected as in
(\ref{24}), their bilinear combination $v^{a'}=\psi_1\mathcal
P_6\gamma^{a'}\mathcal P_6\psi_2=\psi_1^{b'}(\mathcal
P_6\gamma^{a'}\mathcal P_6)_{b'c'}\psi_2^{c'}$ transforms as a
6--dimensional 'vector'.

\def\theequation{B.\arabic{equation}}\label{B}
\section{Appendix B. $OSp(6|4)/U(3)\times SO(1,3)$ supercoset realization
and other ingredients of the $(10|32)$--dimensional $AdS_4\times
CP^3$ superspace}
\setcounter{equation}0

The supervielbeins and the superconnections of the
$OSp(6|4)/U(3)\times SO(1,3)$ supercoset which appear in the
definition of the geometric and gauge quantities of the $AdS_4\times
CP^3$ superspace in Section \ref{superspace} are
\be\label{cartan24}
\begin{aligned}
E^{\hat a}&=e^{\hat a}(x)+4i\vartheta\gamma^{\hat
a}\,{{\sinh^2{{\mathcal M}_{24}/ 2}}\over{\mathcal M}^2_{24}}\,
D_{24}\vartheta,\\
E^{a'}&=e^{a'}(y)+4i\vartheta\gamma^{a'}\gamma^5\,{{\sinh^2{{\mathcal
M}_{24}/2}}\over{\mathcal M}_{24}^2}\,D_{24}\vartheta\,,
\\
E^{\alpha a'}&=\left({{\sinh{\mathcal M}_{24}}\over{\mathcal
M}_{24}}D_{24}\vartheta\right)^{\alpha a'},\\
\Omega^{\hat a\hat b}&=\omega^{\hat a\hat b}(x)+\frac{8}{R}\vartheta\gamma^{\hat a\hat b}\gamma^5\,
{{\sinh^2{{\mathcal M}_{24}/2}}\over{\mathcal M}_{24}^2}D_{24}\vartheta\,,\\
\Omega^{a'b'}&=\omega^{a'b'}(y)-\frac{4}{R}
\vartheta(\gamma^{a'b'}-iJ^{a'b'}\gamma^7)\gamma^5\,{{\sinh^2{{\mathcal M}_{24}/2}}
\over{\mathcal M}_{24}^2}\,D_{24}\vartheta\,,\\
A&=\frac{1}{8}J_{a'b'}\Omega^{a'b'}=A(y)+\frac{4i}{R}\,\vartheta\gamma^7\gamma^5\,{{\sinh^2{{\mathcal
M}_{24}/2}}\over{\mathcal M}_{24}^2}\,D_{24}\vartheta\,,
\end{aligned}
\ee
where
\be\label{M24}
R\,({\mathcal M}_{24}^2)^{\alpha a'}{}_{\beta b'}=
4\vartheta^{\alpha}_ {b'}\,(\vartheta^{a'}\gamma^5)_\beta
-4\delta^{a'}_{b'}\vartheta^{\alpha c'}(\vartheta\gamma^5)_{\beta
c'} -2(\gamma^5\gamma^{\hat a}\vartheta)^{\alpha
a'}(\vartheta\gamma_{\hat a})_{\beta b'} -(\gamma^{\hat a\hat b}\vartheta)^{\alpha
a'}(\vartheta\gamma_{\hat a\hat b}\gamma^5)_{\beta b'}\,.
\ee
The derivative appearing in the above equations is defined as
\be\label{D24}
D_{24}\vartheta={\mathcal P_6}\,(d +\frac{i}{R}\,e^{\hat
a}\gamma^5\gamma_{\hat a}
+\frac{i}{R}e^{a'}\gamma_{a'}-\frac{1}{4}\omega^{\hat a\hat
b}\gamma_{\hat a\hat b}
-\frac{1}{4}\omega^{a'b'}\gamma_{a'b'})\vartheta\,,
\ee
where $e^{\hat a}(x)$, $e^{a'}(y)$, $\omega^{\hat a\hat b}(x)$, $\omega^{a'b'}(y)$
and $A(y)$ are the vielbeins and connections of the bosonic
solution. The $U(3)$--connection $\Omega^{a'b'}$ satisfies the
condition
\begin{equation}
{(P^{-})_{a'b'}}^{c'd'}\Omega_{c'd'}=\frac{1}{2}\,({\delta_{[a'}}^{c'}\,{\delta_{b']}}^{d'}\,-\,
{J_{[a'}}^{c'}\,{J_{b']}}^{d'})\Omega_{c'd'}=0\,,
\end{equation}
where $J_{a'b'}$ is the K\"ahler form on $CP^3$.

\def\thesubsection{B.1}
\subsection{Other quantities appearing in the definition of the
$AdS_4\times CP^3$ superspace of Section \ref{superspace}}

\be\label{M}
R\,({\mathcal M}^2)^{\alpha i}{}_{\beta j}= 4(\ve\upsilon)^{\alpha
i}(\ups\ve\gamma^5)_{\beta j} -2(\gamma^5\gamma^{\hat
a}\upsilon)^{\alpha i}(\ups\gamma_{\hat a})_{\beta j} -(\gamma^{\hat
a\hat b}\upsilon)^{\alpha i}(\ups\gamma_{\hat a\hat
b}\gamma^5)_{\beta j}\,,
\ee
\be
(m^2)^{ij}=-\frac{4}{R}\upsilon^i\,\gamma^5\,\upsilon^j\,,
\ee

\be
\begin{aligned}
\Lambda_{\hat a}{}^{\hat b}&=
\delta_{\hat a}{}^{\hat b}-\frac{R^2}{k^2l_p^2}\,\cdot\,
\frac{e^{-\frac{2}{3}\phi}}{e^{\frac{2}{3}\phi}
+{R\over{kl_p}}\,\Phi}\,{E_{7\hat a}}\,E_7{}^{\hat b}\,,
\\
\\
S_{\underline\beta}{}^{\underline\alpha}&=
\frac{e^{-\frac{1}{3}\phi}}{\sqrt2}\left(\sqrt{e^{\frac{2}{3}\phi}
+{R\over{kl_p}}\,\Phi}-{R\over{kl_p}}\,
\frac{E_7{}^{\hat a}\,\Gamma_{\hat a}\Gamma_{11}}{\sqrt{e^{\frac{2}{3}\phi}
+{R\over{kl_p}}\,\Phi}}
\,\right)_{\underline\beta}{}^{\underline\alpha}
\end{aligned}
\ee

\be\label{phiE7}\begin{aligned}
E_7{}^{\hat a}(\upsilon)&=-\frac{8i}{R}\,\upsilon\gamma^{\hat a}\,{{\sinh^2{{\mathcal
M}/ 2}}\over{\mathcal M}^2}\,\varepsilon\,{\upsilon}\,,
\\
\Phi(\upsilon)&= 1+\frac{8}{R}\,\upsilon\,\ve\gamma^5\,{{\sinh^2{{\mathcal
M}/2}}\over{\mathcal M}^2}\,\ve\upsilon\,.
\end{aligned}
\ee
Let us emphasise that the $SO(2)$ indices $i,j=1,2$ are raised and
lowered with the unit matrices $\delta^{ij}$ and $\delta_{ij}$ so
that there is actually no difference between the upper and the
lower $SO(2)$ indices, $\varepsilon_{ij}=-\varepsilon_{ji}$,
$\varepsilon^{ij}=-\varepsilon^{ji}$ and
$\varepsilon^{12}=\varepsilon_{12}=1$.

\def\theequation{C.\arabic{equation}}\label{C}
\section{Appendix C. Identities for the kappa-projected fermions}
\setcounter{equation}0

When the fermionic variables
$\Theta^{\underline\alpha}=(\vartheta^{\alpha a'},\,\upsilon^{\alpha
i})$ are subject to the constraint (\ref{kappagauge1}), the
following identities hold.

\def\thesubsection{C.1}
\subsection{Identities involving $\upsilon^{\alpha i}$}
\be\label{i1}
\ups^i\gamma^5\ups^j=\ups^i\gamma^3\ups^j=0\,,\qquad \ups^{\alpha i}\ups^{\beta j}\delta_{ij}=-\frac{1}{4}((1\pm\gamma)C^{-1})^{\alpha\beta}\ups\ups\,,
\ee
where $\gamma=\gamma^{012}$ and $\ups\ups=\delta_{ij}\ups^{\alpha
i}C_{\alpha\beta}\ups^{\beta j}$.

Another useful relation is ($\ve^{012}=-\ve_{012}=1$)
\be\label{gg}
\ups\gamma_{ab}d\ups=\pm\ve_{abc}\ups\gamma^cd\ups\,,
\ee
which also holds for the kappa--projected $\vartheta$ and
$d\vartheta$.

Using eqs. (\ref{i1}) and (\ref{gg}) we find that
\be
\ups\ve\gamma^a\ups\,\ups\ve\gamma_b\ups=\delta_b^a(\ups\ups)^2\,,
\qquad \ups\ve\gamma^{ac}\ups\,\ups\ve\gamma_{cb}\ups=2\delta_b^a(\ups\ups)^2\,,
\ee
\be
(m^2)^{ij}=-\frac{4}{R}\upsilon^i\,\gamma^5\,\upsilon^j=0
\ee
and
\bee
({\mathcal M}^2\ve\ups)^{\alpha i}=0\,.
\eee
 A similar computation shows that
\be
\ups\ve\gamma^5{\mathcal M}^2=0.
\ee
It is also true in general (i.e. without fixing $\kappa$--symmetry)
that
\be
{\mathcal M}^2\ups=0\,,\qquad \ups\gamma^5{\mathcal M}^2=0.
\ee
Using the above identities we find that for $\upsilon$ satisfying
(\ref{kappagauge1})
\be\label{M2d}
\mathcal M^2D\upsilon
=\frac{6i}{R^2}(E^a\pm\frac{R}{2}\Omega^{a3})(\gamma_a\ups)\,\ups\ups
\ee
which results in
\be\label{vmdv}
4\upsilon\gamma^a \frac{\sinh^2(\mathcal M/2)}{\mathcal
M^2}D\upsilon=
\upsilon\gamma^a(1+\frac{1}{12}\mathcal M^2)D\upsilon
=\upsilon\gamma^a\,(d-\frac{1}{4}\Omega^{bc}\gamma_{bc})\upsilon
+\frac{i}{2R^2}(E^a\pm\frac{R}{2}\Omega^{a3})(\ups\ups)^2\,,
\ee
where $E^a$, $\Omega^{bc}$ and $\Omega^{a3}$ are $AdS_4$ components
of the supervielbein and connection of the supercoset
$OSp(6|4)/U(3)\times SO(1,3)$ defined in eqs. (\ref{cartan24}) and
the matrix ${\mathcal M}^2$ is defined in eq. (\ref{M}).

We also find that
\be
4\ups\ve\gamma^5{{\sinh^2{{\mathcal M}/2}}\over{\mathcal M}^2}D\ups
=\ups\ve\gamma^5D\ups=\frac{i}{R}(E^a\pm\frac{R}{2}\Omega^{a3})\ups\ve\gamma_a\ups\,.
\ee

\def\thesubsection{C.2}
\subsection{Identities involving $\vartheta^{\alpha a'}$ and the simplified
form of the \\ $OSp(6|4)/U(3)\times SO(1,3)$ supergeometry}

Using the definition of $\mathcal M_{_{24}}$, eq. (\ref{M24}), and
the fact that
\be
[\gamma^{012},\gamma^{a'}]=0
\ee
we find that
\be
(\vartheta\gamma'\gamma^5{\mathcal M}_{_{24}}^2)_{\beta b'} =0
\qquad
({\mathcal M}_{_{24}}^2\gamma'\vartheta)^{\alpha a'}=0\,,
\ee
where $\gamma'$ is any product of the gamma-matrices that commutes
with $\gamma=\gamma^{012}$,
\emph{e.g.} any product of $\gamma^{a'}$  and $\gamma^{a}$. A
slightly longer computation, using the fact that
\be
\gamma^3\vartheta=\pm\gamma^3\gamma^{012}\vartheta=\pm i\gamma^5\vartheta\,,\qquad\vartheta\gamma^3=\mp i\vartheta\gamma^5\qquad\mbox{for}\quad\vartheta=\frac{1}{2}(1\pm\gamma)\vartheta\,,
\ee
shows that with this projection of the $\vartheta$s
\be
\mathcal M_{_{24}}^4=0\,.
\ee
Using the identity
\be
\vartheta^{\alpha a'}\vartheta^{\beta b'}\delta_{a'b'}=-\frac{1}{4}((1\pm\gamma)C^{-1})^{\alpha\beta}\vartheta\vartheta\,,
\ee
where $\vartheta\vartheta\equiv\vartheta^{\alpha
a'}C_{\alpha\beta}\vartheta^{\beta b'}\,\delta_{a'b'}\,,$
 one can further show that
\be
({\mathcal M}^2_{_{24}}D_{_{24}}\vartheta)^{\alpha a'}
=\frac{6i}{R^2}(e^b\pm\frac{R}{2}\omega^{b3})(\gamma_b\vartheta)^{\alpha
a'}\,\vartheta\vartheta,
\ee
where the covariant derivative $D_{24}$, defined in
(\ref{D24}), becomes
\be
D_{24}\vartheta=\mathcal P_6(d
+\frac{i}{R}(e^a\pm\frac{R}{2}\omega^{a3})\gamma^5\gamma_a
\mp\frac{1}{R}\,e^3
+\frac{i}{R}e^{a'}\gamma_{a'}
-\frac{1}{4}\omega^{ab}\gamma_{ab}
-\frac{1}{4}\omega^{a'b'}\gamma_{a'b'})\vartheta\,.
\ee
This gives
\bee
\vartheta\gamma^a(1+\frac{1}{12}{\mathcal M}^2_{_{24}})D_{_{24}}\vartheta
&=&
\vartheta\gamma^aD_{24}\vartheta+\frac{i}{2R^2}(e^a\pm\frac{R}{2}\omega^{a3})(\vartheta\vartheta)^2\,.
\eee
Using the above expressions one finds that the form of the
$OSp(6|4)/U(3)\times SO(1,3)$ geometrical objects (\ref{cartan24})
simplify to
\be
\begin{aligned}
E^a&=e^a(x)+i\vartheta\gamma^aD_{24}\vartheta-\frac{1}{2R^2}(e^a\pm\frac{R}{2}\omega^{a3})(\vartheta\vartheta)^2,
\\
E^3&=e^3(x),\\
E^{a'}&=e^{a'}(y)-\frac{1}{R}(e^a\pm\frac{R}{2}\omega^{a3})\vartheta\gamma^{a'}\gamma_a\vartheta
\\
E^{\alpha a'}&=
\left(D_{_{24}}\vartheta\right)^{\alpha a'}
+\frac{i}{R^2}(e^b\pm\frac{R}{2}\omega^{b3})(\gamma_b\vartheta)^{\alpha
a'}\,\vartheta\vartheta,
\\
\Omega^{ab}&=\omega^{ab}(x)+\frac{2i}{R^2}(e^c\pm\frac{R}{2}\omega^{c3})\vartheta\gamma^{ab}{}_c\vartheta,
\\
\Omega^{a3}&=\omega^{a3}(x)
\mp\frac{2i}{R}\vartheta\gamma^aD_{24}\vartheta
\pm\frac{1}{R^3}(e^a\pm\frac{R}{2}\omega^{a3})(\vartheta\vartheta)^2,
\\
\Omega^{a'b'}&=\omega^{a'b'}(y)-\frac{i}{R^2}(e^a\pm\frac{R}{2}\omega^{a3})\vartheta(\gamma^{a'b'}-iJ^{a'b'}\gamma^7)\gamma_a\vartheta,
\\
A&=A(y)-\frac{1}{R^2}(e^a\pm\frac{R}{2}\omega^{a3})\vartheta\gamma^7\gamma_a\vartheta
\,,
\end{aligned}
\ee
and in particular
\be
E^a\pm\frac{R}{2}\Omega^{a3}=e^a(x)\pm\frac{R}{2}\omega^{a3}(x)\,.
\ee
Thus, in the chosen $\kappa$--symmetry gauge the
$OSp(6|4)/U(3)\times SO(1,3)$ supercoset geometry depends on the
fermionic coordinates only up to the 4th power.

Note that in all the above expressions the components $e^a(x)$
$(a=0,1,2)$ and ${R/2}\,\omega^{a3}(x)$ of the $AdS_4$ vielbein and
connection appear only in the combination $e^a(x)\pm
{R/2}\,\omega^{a3}(x)$. This combination has a very clear
geometrical meaning. In the case, when the indices $a=0,1,2$ label
the directions of the 3d Minkowski slice of the $AdS_4$, $e^a(x)\pm
{R/2}\,\omega^{a3}(x)$ corresponds to the generator
$\Pi_a=P_a\mp{1/2}\,M_{a3}$ of the Poincar\'e translations
($[\Pi_a,\Pi_b]=0$) along the 3d Minkowski boundary which is the
linear combination of boosts and Lorentz rotations in $AdS_4$ (see
\cite{Pasti:1998tc} for more details). More precisely, $e^a(x)-
{R/2}\,\omega^{a3}(x)$ corresponds to the Poincar\'e translation,
while $e^a(x)+{R/2}\,\omega^{a3}(x)$ corresponds to the conformal
boosts in $M_3$, or vice versa, depending on the orientation.

When the $AdS_4$ metric is chosen in the form (\ref{ads4metric11})
the vielbein $e^a(x)$ and the connection $\omega^{a3}(x)$ are
proportional to each other, namely,
\be\label{eaoa3}
e^a=-{R\over 2}\,\omega^{a3}.
\ee
Actually, this relation can be imposed for any
form of the metric by performing an appropriate $SO(1,3)$
transformation of the $AdS_4$ vielbein and connection.

\end{document}